\documentclass[preprint]{aastex}

\usepackage{natbib}
\newcommand{\tnos}{\textsc{tno}s}
\newcommand{\tno}{\textsc{tno}}
\newcommand{\ossos}{\textsc{ossos}\ }
\newcommand{\cfeps}{\textsc{cfeps}\ }
\newcommand{\au}{\textsc{au}\ }

\shorttitle{OSSOS: design and first discoveries}
\shortauthors{Bannister et al.}

\begin{document}

\title{The Outer Solar System Origins Survey: I. Design and First-Quarter Discoveries}

\author{Michele T. Bannister\altaffilmark{1,2}, 
J. J. Kavelaars\altaffilmark{1,2}, 
Jean-Marc Petit\altaffilmark{3}, 
Brett J. Gladman\altaffilmark{4}, 
Stephen D. J. Gwyn\altaffilmark{2}, 
Ying-Tung Chen\altaffilmark{5}, 
Kathryn Volk\altaffilmark{6}, 
Mike Alexandersen\altaffilmark{4, 5}, 
Susan Benecchi\altaffilmark{7},
Audrey Delsanti\altaffilmark{8}, 
Wesley Fraser\altaffilmark{2, 9}, 
Mikael Granvik\altaffilmark{10, 11}, 
Will M. Grundy\altaffilmark{12}, 
Aur\'elie Guilbert-Lepoutre\altaffilmark{3}, 
Daniel Hestroffer\altaffilmark{13}, 
Wing-Huen Ip\altaffilmark{14, 15},
Marian Jakubik\altaffilmark{16}, 
Lynne Jones\altaffilmark{17}, 
Nathan Kaib\altaffilmark{18},
Catherine F. Kavelaars\altaffilmark{1},
Pedro Lacerda\altaffilmark{19}, 
Samantha Lawler\altaffilmark{2},
Matthew J. Lehner\altaffilmark{5, 20, 21}, 
Hsing Wen Lin\altaffilmark{14}, 
Tim Lister\altaffilmark{22}, 
Patryk Sofia Lykawka\altaffilmark{23}, 
Stephanie Monty\altaffilmark{1},
Michael Marsset\altaffilmark{8, 24},
Ruth Murray-Clay\altaffilmark{25}, 
Keith Noll\altaffilmark{26}, 
Alex Parker\altaffilmark{27},
Rosemary E. Pike\altaffilmark{1}, 
Philippe Rousselot\altaffilmark{3}, 
David Rusk\altaffilmark{1},
Megan E. Schwamb\altaffilmark{5},
Cory Shankman\altaffilmark{1}, 
Bruno Sicardy\altaffilmark{28}, 
Pierre Vernazza\altaffilmark{8}, 
Shiang-Yu Wang\altaffilmark{5}
}
\email{micheleb@uvic.ca}
\altaffiltext{1}{Department of Physics and Astronomy, University of Victoria, Elliott Building, 3800 Finnerty Rd, Victoria, BC V8P 5C2, Canada}
\altaffiltext{2}{NRC-Herzberg Astronomy and Astrophysics, National Research Council of Canada, 5071 West Saanich Rd, Victoria, British Columbia V9E 2E7, Canada}
\altaffiltext{3}{Institut UTINAM UMR6213, CNRS, Univ. Bourgogne Franche-Comt\'e, OSU Theta F25000 Besan\c{c}on, France}
\altaffiltext{4}{Department of Physics and Astronomy, University of British Columbia, Vancouver, BC, Canada}
\altaffiltext{5}{Institute for Astronomy and Astrophysics, Academia Sinica; 11F AS/NTU, National Taiwan University, 1 Roosevelt Rd., Sec. 4, Taipei 10617, Taiwan}
\altaffiltext{6}{Department of Planetary Sciences/Lunar and Planetary Laboratory, University of Arizona, 1629 E University Blvd, Tucson, AZ 85721, United States}
\altaffiltext{7}{Planetary Science Institute, 1700 East Fort Lowell, Suite 106, Tucson, AZ 85719, United States}
\altaffiltext{8}{Aix Marseille Universit\'e, CNRS, LAM (Laboratoire d'Astrophysique de Marseille) UMR 7326, 13388, Marseille, France}
\altaffiltext{9}{Queen's University Belfast, Astrophysics Research Centre, Belfast, Northern Ireland}
\altaffiltext{10}{Department of Physics, P.O. Box 64, 00014 University of Helsinki, Finland}
\altaffiltext{11}{Finnish Geospatial Research Institute, P.O. Box 15, 02430 Masala, Finland}
\altaffiltext{12}{Lowell Observatory, Flagstaff, Arizona, United States}
\altaffiltext{13}{IMCCE, Observatoire de Paris - PSL research univ., UPMC univ. P06, univ. Lille 1, CNRS, F-75014 Paris, France}
\altaffiltext{14}{Institute of Astronomy, National Central University, Taiwan}
\altaffiltext{15}{Space Science Institute, Macau University of Science and Technology, Macau}
\altaffiltext{16}{Astronomical Institute, Slovak Academy of Science, 05960 Tatranska Lomnica, Slovakia}
\altaffiltext{17}{University of Washington, Washington, United States}
\altaffiltext{18}{HL Dodge Department of Physics \& Astronomy, University of Oklahoma, Norman, OK 73019, United States}
\altaffiltext{19}{Max Planck Institute for Solar System Research, Justus-von-Liebig-Weg 3, 37077 G\"ottingen, Germany}
\altaffiltext{20}{Department of Physics and Astronomy, University of Pennsylvania, 209 S. 33rd St., Philadelphia, PA 19104, USA}
\altaffiltext{21}{Harvard-Smithsonian Center for Astrophysics, 60 Garden St., Cambridge, MA 02138, USA}
\altaffiltext{22}{Las Cumbres Observatory Global Telescope Network, Inc., 6740 Cortona Drive Suite 102, Goleta, CA  93117, United States}
\altaffiltext{23}{Astronomy Group, School of Interdisciplinary Social and Human Sciences, Kinki University, Japan}
\altaffiltext{24}{European Southern Observatory (ESO), Alonso de C\'ordova 3107, 1900 Casilla Vitacura, Santiago, Chile}
\altaffiltext{25}{Department of Physics, University of California, Santa Barbara, CA 93106, USA}
\altaffiltext{26}{NASA Goddard Space Flight Center, Code 693, Greenbelt, MD 20771, United States}
\altaffiltext{27}{Southwest Research Institute, Boulder, Colorado, United States}
\altaffiltext{28}{LESIA, Observatoire de Paris, CNRS UMR 8109, Universit\'e Pierre et Marie Curie, Universit\'e Paris-Diderot, 5 place Jules Janssen, F-92195 Meudon Cedex, France}

\begin{abstract}

We report the discovery, tracking and detection circumstances for 85 trans-Neptunian objects (\tnos) from the first 42 deg$^{2}$ of the Outer Solar System Origins Survey (\textsc{ossos}).
This ongoing $r$-band Solar System survey uses the 0.9 deg$^{2}$ field-of-view MegaPrime camera on the 3.6 m Canada-France-Hawaii Telescope.
Our orbital elements for these \tnos\ are precise to a fractional semi-major axis uncertainty $<0.1\%$.
We achieve this precision in just two oppositions, as compared to the normal 3--5 oppositions, via a dense observing cadence 
and innovative astrometric technique.
These discoveries are free of ephemeris bias, a first for large trans-Neptunian surveys.
We also provide the necessary information to enable models of  \tno\ orbital distributions to be tested against our \tno\ sample. 
We confirm the existence of a cold ``kernel'' of objects within the main cold classical Kuiper belt, 
and infer the existence of an extension of the ``stirred'' cold classical Kuiper belt to at least several \au
beyond the 2:1 mean motion resonance with Neptune.
We find that the population model of \citet{Petit:2011p3938} remains a plausible representation of the Kuiper belt.
The full survey, to be completed in 2017, will provide an exquisitely characterized sample of important resonant \tno\ populations,
ideal for testing models of giant planet migration during the early history of the Solar System.


\end{abstract}


\section{Introduction}
We present here the design and initial observations and discoveries of the Outer Solar System Origins Survey (\textsc{ossos}).
\ossos will provide a flux-limited sample of approximately five hundred trans-Neptunian objects (\tnos),
with high-precision, dynamically classified orbits.
The survey is especially sensitive to \tnos\ that are in exterior mean-motion resonance with Neptune. 
\ossos will measure the absolute abundance and orbital distributions of numerous resonant populations,
the main classical belt, the scattering and detached populations,
and the libration amplitude distribution in many low-order resonances.
The \ossos dataset will provide direct constraints on cosmogonic scenarios that attempt to explain the formation of the trans-Neptunian populations. 

Scenarios for the formation of the trans-Neptunian orbital distribution have distinct fingerprints.
Discerning the features of the populations has required many sky surveys; \citet{Bannister:o9a-KucX} reviews these.
The present \tno\ orbital distribution is a signature of excitation events that occurred earlier in the dynamical history of the Solar System \citep{1984Icar...58..109F}. 
Certain features of the orbital distribution are diagnostic of the evolutionary processes that sculpted the disk.
Foremost among these features are the \tnos\ trapped in the mean-motion resonances with Neptune. 
The population abundances and orbital distribution in each mean-motion resonance with Neptune are dependent on the mechanism that emplaced the \tnos\ into resonance.
Proposed mechanisms for the trapping of \tnos\ into resonances include scenarios where objects on low-eccentricity orbits were trapped and pumped to higher eccentricities during subsequent migration \citep[e.g.][]{Malhotra:1995dy,1999AJ....117.3041H,2005AJ....130.2392H}. Alternate scenarios have objects trapped into the resonances out of a scattering population, after which their eccentricities were damped \citep[e.g.][]{2008Icar..196..258L}. 
The \ossos dataset will enable testing of the veracity of proposed models of initial radial planetesimal distribution, planet migration distances and time scales. 
One example of how these scenarios can be tested is by measurement of the present distribution of \tnos\ within the substructure of the 2:1 resonance. The speed of Neptune's past migration influences the present ratio of objects leading or trailing Neptune in orbital longitude \citep{MurrayClay:2004hh}.  However, the population asymmetry appears to be small; the discovery of more \tnos\ that orbit within these diagnostic features is therefore required \citep[e.g.][]{MurrayClay:2004hh, Gladman:2012ed}. 
\ossos will provide sufficient \tno\ orbits to precisely measure the distinct fingerprints of these alternative formation scenarios.

Distant n:1 and n:2 resonances with semi-major axes above 50 \au harbour significant stable populations 
formed during the early history of the Solar System.
\citet{Chiang:2003hb} and \citet{Elliot:2005ju} were the first to report objects in the 5:2 and 7:3 resonances, 
while \citet{Lykawka:2007kc} and \citet{Gladman2008} reported in detail resonant TNOs in several distant resonances. 
Later, \citet{Gladman:2012ed} characterized the main properties of those distant populations.
More recently \citet{Pike:2015gn} found evidence for a substantial population in the distant 5:1 resonance, 
rivalling in number the closer 3:2 resonant population.
Assessing the intrinsic populations and eccentricity/inclination/libration amplitude distributions of the populations in distant resonances 
will help clarify if temporary resonance trapping by scattering TNOs (resonance sticking) 
or capture during planet migration played the major role in producing those populations \citep{Chiang:2003hb, Lykawka:2007ff}.
These outer resonances constrain both the mechanisms that operated at that time (e.g. the behaviour of planetary migration), 
and the orbital properties and extent of the protoplanetary disk \citep{2005AJ....130.2392H, Lykawka:2007ff}.
\ossos has sensitivity to the distant resonances and will unveil unprecedented details of the resonant structure beyond 50 \textsc{au}.

\ossos is designed to discover the necessary new sample of \tnos\ in a way that allows the underlying populations' orbit distribution to be determined.
\tno\ discovery is inherently prone to observationally induced biases \citep{Trujillo:2000hm, Jones:2006jl, Kavelaars:ws, Jones:2010bb}.
To be detected, an object has to be brighter than a survey's flux limit, while moving within the area of sky that the survey is examining.
Resonant \tnos\ can have highly eccentric orbits ($e \gtrsim 0.1$) that explore large heliocentric distances where they become too faint to detect: \tnos\ are brightest at their pericenter.
Owing to the steep \tno\ size distribution \citep{Fraser:2009im}, most \tnos\ detected in a given survey will be small and near the flux limit.
For example, the 5:1 resonance has such a large semi-major axis (88 \textsc{au}) that a typical object in the 5:1 would
only be visible in a flux-limited sample for $<1$\% of its orbital period \citep{Gladman:2012ed}.
Minimal loss of objects following their discovery, and accurate survey debiasing, are necessary to ascertain the population structure of these hard-to-sample resonances. 

\ossos builds on the experiences and lessons of data acquisition from the more than sixty discovery surveys in past decades (listed in \citet{Bannister:o9a-KucX}) that have brought us to our current understanding of the trans-Neptunian region.
Crucially, we aim to acquire a \tno\ sample free from the challenging problem of \textit{ephemeris bias} \citep{Jones:2006jl}: selection effects due to choices of orbit estimation and of recovery observations.
\ossos is conducted as a queue-mode Large Program with the MegaCam imager on the 3.6 m Canada-France-Hawaii Telescope (CFHT) to discover and to follow-up our discoveries. Follow-up is $>90\%$ of the survey's 560-hour time budget, and allows us to constrain the orbits of our discoveries with exquisite precision. This removes the need for follow-up to confirm orbits by facilities other than the survey telescope. Objects are tracked until their orbital classification (\S~\ref{sec:classification}) is secure, which at minimum requires reaching semi-major axis uncertainties $\sigma_{a} < 1$\%, and which may require reaching $\sigma_{a} < 0.01$\%.
We describe here our observation strategy, our astrometric and photometric calibration, the open-source data processing pipeline, the characterization of our \tno\ detection efficiency, the survey's simulator, and the discoveries in the first quarter of the survey.

\section{Survey design and observations}
\label{sec:design}

\begin{figure}
\plotone{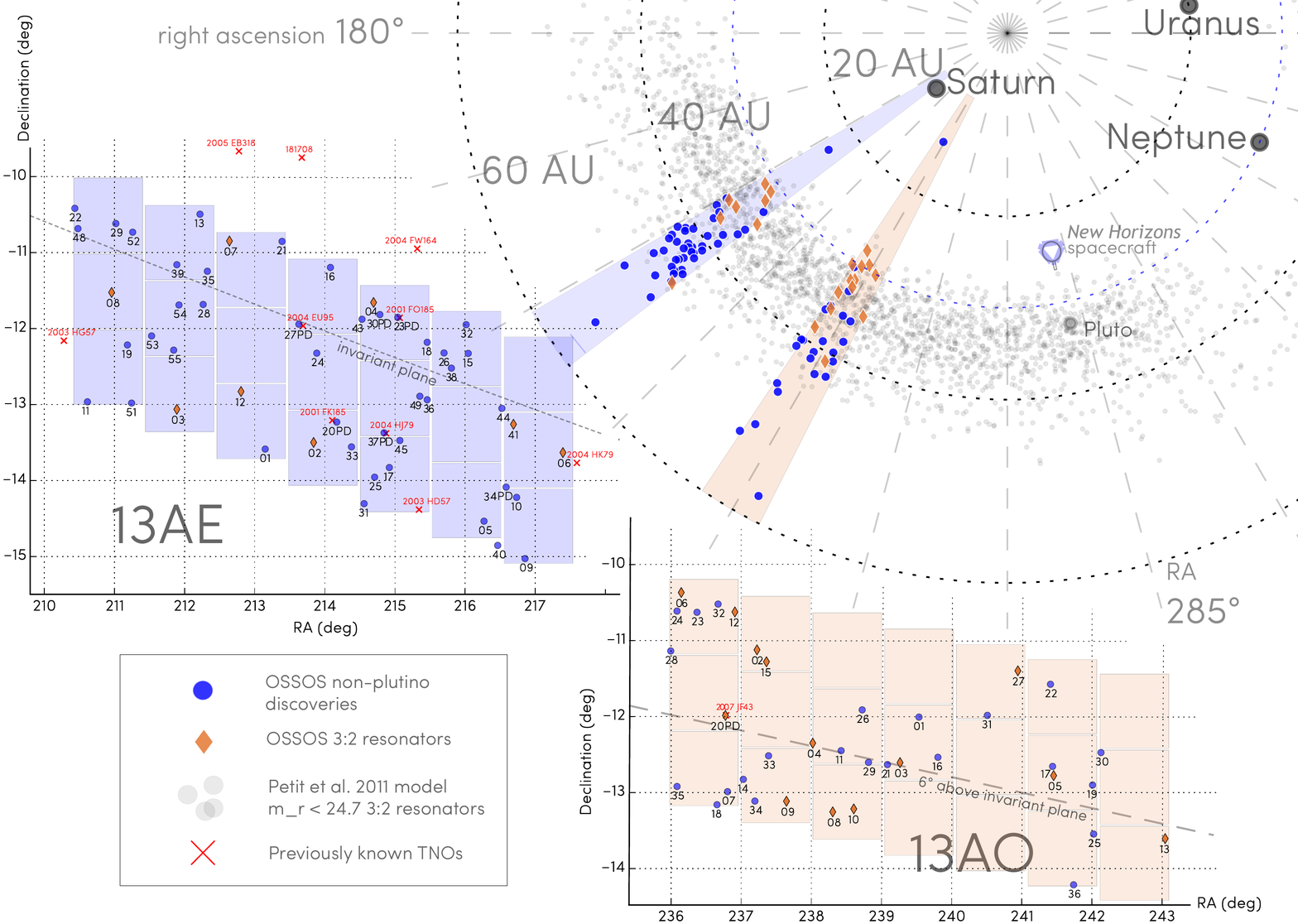}
\caption{The first-quarter survey coverage for \ossos relative both to the geometry of the Solar System (top right) and to the sky (top left and bottom right), at the time of the discovery observations in 2013 (blue: 13AE, orange: 13AO). Characterised discoveries (\S~\ref{sec:characterization}) are labelled with the last two digits of Table~\ref{tab:discoveries} for their respective blocks. The on-plane 13AE block contains more discoveries (49 \tnos) than in the higher-latitude 13AO block (36 \tnos), due to the cold classical Kuiper belt's concentration in the plane.
The grey background points in the top right view show a prediction of the position density of Plutinos (objects in the 3:2 resonance with Neptune) with instantaneous $m_{r} < 24.7$, as modelled by \citet{Gladman:2012ed}.
Plutinos avoid the longitude of Neptune due to the resonance's protection mechanism.
The visible model population is biased by detection proximate to perihelion. Plutinos discovered by \ossos (orange diamonds) display this perihelion bias; note that 13AO is is close to the location where Plutinos with zero libration amplitude are currently coming to perihelion.
\label{fig:13Ablocks}}
\end{figure}

The \ossos observations are acquired in \textit{blocks:} contiguous patches of sky formed by a layout of adjoining multiple 0.90 deg$^{2}$ MegaCam fields. These are made large enough to reduce the chance of losing objects due to orbit shear, and sufficiently narrow in right ascension to be easily queue-scheduled for multiple observations in a single night.   
For the discovery blocks reported here (\S~\ref{sec:13A}), a 3 x 7 grid of pointings was used to achieve this goal (Fig.~\ref{fig:13Ablocks}).  

The survey is observed in two parts, as a given right ascension can only be observed for $\sim\! 6$ months at a time. During the discovery opposition, a block is observed multiple times in each of five to six lunations to provide a robust initial estimation of the orbits of discovered objects. Field centers shift during this time by drifting the block over the six months at the Kuiper belt average sky motion rate, which tracks the \tnos\ present in that area of sky (\S~\ref{sec:cadence}).
A year later, the next opposition is dedicated to discovery followup (\S~\ref{sec:cadence}). The orbit determination from
the first year is good enough to allow pointed recoveries of each object during the second year (\S~\ref{sec:linkage}).
The survey cadence was based on simulations of ephemeris sampling under nominal CFHT observing conditions. 
The simulations determined the cadence required to reduce the nominal fractional semi-major axis uncertainty, $\sigma_{a}$, to the level required for secure dynamical orbital classification within two years of first detection. 
The orbital uncertainty reduces in a complex fashion, dependent on total arc length observed, number of observations, time of these observations relative to the opposition point, and an object's heliocentric distance: closer objects benefit from a larger parallactic lever arm due to the Earth's motion. 
The cadence we selected ensured resonant identification was probable in the discovery year, with the second year's observations needed to determine the libration amplitude with reasonable precision.  
A survey of 32 deg$^{2}$ in 2011-12 by \citet{Alexandersen:2014uv} showed that this mode successfully provided classifiable orbits within two years of discovery.

Resonance dynamics require that resonant objects come to pericenter at a set longitude relative to Neptune \citep[e.g.][]{Volk:2015}. This confines the sky locations of their perihelia to a restricted range of ecliptic longitudes. 
Each \ossos block location (listed in Table~\ref{tab:blocks}) was placed at ecliptic longitudes that maximize the detections of objects in certain low-order resonances with Neptune. Table~\ref{tab:blocks} gives only the types of resonances that will have objects with small libration amplitudes that are at perihelion at those locations; resonance sensitivities are a complex function of orbital libration amplitude and eccentricity, and are discussed further in \citet{Volk:2015}.
The exact on-sky block placement is chosen to avoid chip-saturating stars brighter than $m_r = 12$ and \tno-obscuring features like open clusters. 
We also avoid placement near the galactic plane, due to severe stellar crowding in this region. 
Extracting the complex biases that this sky placement causes on the detection of objects from the underlying population \citep{Gladman:2012ed,Lawler:2013hp} is accounted for by the \ossos survey simulator (\S~\ref{sec:simulator}).  

A pair of blocks was observed in each half-year CFHT semester. Each semester's pair was sited to maximize sampling of populations that occupy a range of inclinations. One block targeted the highest density of \tnos. This density centres closer to the invariable plane \citep{Souami:2012fc} than to the ecliptic \citep{Chiang:2008dz, Elliot:2005ju, Brown:2004p96, CollanderBrown:2003ei}. 
The other block was placed between five and ten degrees off the invariable plane. 

\begin{deluxetable}{crrrrlcl}
\tablecolumns{7}
\tablecaption{Target regions for the \ossos survey \label{tab:blocks}}
\tablehead{\colhead{Block} & \colhead{RA} & \colhead{Dec} & \colhead{Ec. lat.}  \vspace{-0.12cm} & \colhead{Angle from} & \colhead{Main resonance} & \colhead{Grid} & \colhead{Observation} \\
\colhead{} & \colhead{($^{\circ}$)} & \colhead{ ($^{\circ}$)} & \colhead{($^{\circ}$)} & \colhead{Neptune ($^{\circ}$)} & \colhead{sensitivities} & \colhead{layout} & \colhead{}
 }
\startdata
15AP 	& 202.5 	& -7.8 	& 1.5 	& -135	& n:2, n:4 		& 4 x 5	& 2015-04 \\ 
13AE 	& 213.9 	& -12.5 	& 1.0		& -119 	& n:2 		& 3 x 7	& 2013-04; this work \\ 
15AM 	& 233.8 	& -12.2 	& 6.9		& -105	& n:2			& 4 x 5	& 2015-05 \\ 
13AO 	& 239.5 	& -12.3 	& 8.0		& -94	& n:2			& 3 x 7	& 2013-05; this work \\ 
15BS 	& 7.5 	& 5.0 	& 1.6 	&  31		& n:1, n:4		& 4 x 5	& 2015-09 \\ 
13BL 	& 13.5 	& 3.8 	& -1.8 	&  41		& n:1, n:3, n:4	& 3 x 7	& 2013-10 \\ 
14BH 	& 22.5 	& 13.0 	& 3.3 	&  51		& n:1, n:3, n:4	& 3 x 7	& 2014-10 \\ 
15BD 	& 48.8 	& 16.5 	& -1.5 	&  74		& n:2, n:3		& 4 x 5	& 2015-11 \\ 
\enddata
\tablecomments{Block names indicate the year (2013-2015) that the discovery observations were successfully made, the half-year semester of discovery opposition (A for Northern spring, B for Northern autumn), and a distinguishing letter. 
Coordinates are the center of each block at the time of discovery when the block reaches opposition.
Angle from Neptune is approximated to projection to the ecliptic at the time of discovery: positive angles lead Neptune, negative angles trail Neptune.
Resonances for each block are only for small libration-amplitude orbits at perihelion. For detailed maps of the 13A blocks' sensitivity to given mean-motion resonances with Neptune, see \citet{Volk:2015}.
In 2015 the configuration of the MegaCam focal plane was altered from 36 to 40 CCDs. 
This required rearranging the tessellation of the fields from the 3 x 7 grid to a 4 x 5 grid.
}
\end{deluxetable}

\subsection{Observing parameters}
\label{sec:parameters}
 
The \ossos discovery and tracking program use the CFHT MegaPrime/MegaCam \citep{2003SPIE.4841...72B}.  
In 2013 and 2014, the MegaPrime/MegaCam focal plane was populated by thirty-six $4612\times2048$ pixel CCDs in a 4 by 9 arrangement, with a $0.96^{\circ} \times 0.94^{\circ}$ unvignetted field of view (0.90 deg$^{2}$)
and 0.05'' full width at half maximum (FWHM) image quality (IQ) variation between centre and edge.
The plate scale is 0.184'' per pixel, which is well suited for sampling the 0.7'' median seeing at Maunakea. 

We observed our 2013 discovery fields in MegaCam's $r$.MP9601 filter (564--685 nm at 50\% transmission; 81.4\% mean transmission), henceforth referred to as $r$, which is similar to the Sloan Digital Sky Survey $r'$ filter (see \S~\ref{sec:photometry}).  Using this filter optimizes the tradeoff between reflected solar brightness [\tnos\ have colours $B-R \sim\! 1-2$ \citep{Hainaut:2012jb}], the telescope's and CCDs' combined quantum efficiency curve, and sky brightness.
The $r$ band delivers the best IQ distribution at CFHT and minimizes IQ distortion from atmospheric dispersion, especially useful as tracking observations often occur months from opposition when the airmass is $>1.3$. 
Obtaining all discovery observations using the same filter simplifies the design of the survey's simulator (\S~\ref{sec:characterization}) and avoids object-color based biases in tracking. 

Our integration length was set at 287~s. 
This exposure length achieves a target depth of $m_{r} = 24.5$ in a single frame in 0.7" median CFHT seeing. 
It reduces loss of signal-to-noise (SNR) due to trailing, with motion during the exposure of less than half a FWHM for objects at $d \geq 33$ AU, also aiding the detection of \tno\ binarity. 
The number of fields in a block is set by the requirement of being able to observe one-half of a block three times (three observations provide the minimal initial orbital constraints for discovery) in three hours, the maximum time over which both airmass and IQ stability can typically be maintained. 
 Given the 40 s MegaCam readout overhead on top of the integration time, this requirement lets us set a grid of approximately 20 fields per block, with the exact number set to give a symmetric grid.
 The survey target depth allows detection of Plutinos with radii larger than 20~km at their perihelion [per \citet{Luu:1988p2993}; assuming a 10\% albedo per \citet{Mommert:2012wj, Peixinho:2015bw}], potentially examining the size distribution where models \citep{Kenyon:2008ky, Fraser:2009wq} and observations \citep{Bernstein:2004p75, Fraser:2009im, Fuentes:2009db, Shankman:2012vh, Alexandersen:2014uv, Fraser:2014vt} suggest a transition in the size distribution.

MegaPrime/MegaCam operates exclusively as a dark-time queue-mode instrument for CFHT. 
The \ossos project thus has between ten and fourteen potentially observable nights each month for observations, weather considerations aside. 
Through CFHT's flexible queue-schedule system we requested our observations be made in possibly non-photometric conditions (discussed in \S~\ref{sec:photometry}) with 0.6--0.8'' seeing and $<0.1$ magnitudes extinction for discovery, and requested image quality of 0.8--1.0'' seeing for followup observations. Images were taken entirely with sidereal guiding and above airmass 1.5. This aided the quality of the astrometric solution and the point spread function, and retained image depth: median extinction on Maunakea is 0.10 mags per airmass in this passband \citep{TheNearbySuperNovaFactory:2012gv}.

\subsection{Cadence}
\label{sec:cadence}

\begin{figure}
\plotone{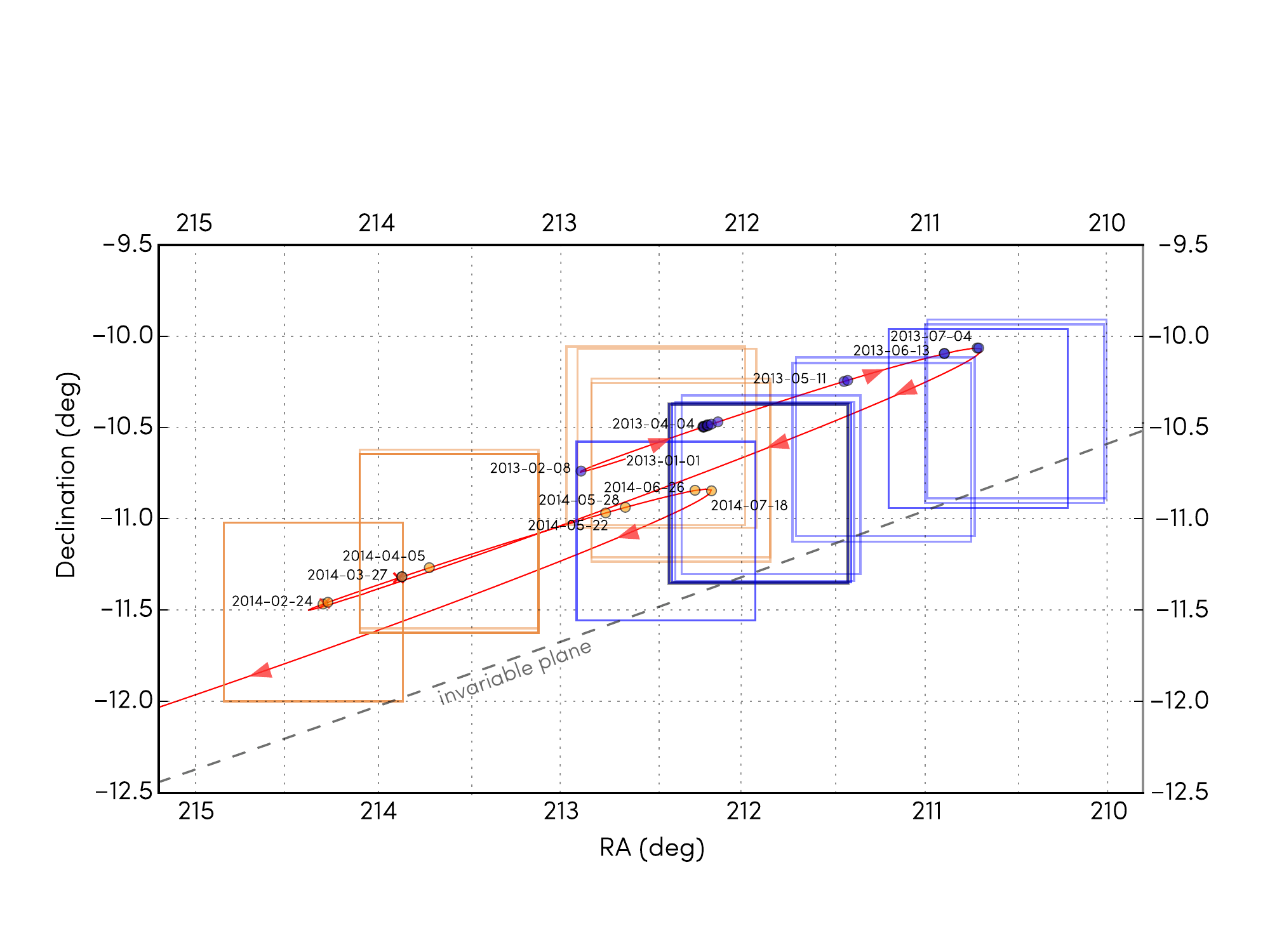}
\caption{The cadence over 2013-2014 of \ossos observations of a single 13AE field and the resultant tracking of one of the \tnos\ discovered in that field, the insecurely (\S~\ref{sec:classification}) resonant object \texttt{o3e13} (Table~\ref{tab:discoveries}). Each box is an exposure of a 36-CCD MegaPrime square field of view. In 2013, the field centre was shifted at the Kuiper belt's average sky motion rate (blue boxes). Note the dense observing cadence during the discovery opposition in April 2013 (heavier blue box due to overlap): the triplet of observations used for object discovery is on April 4, with other imaging April 5 and 6.  In 2014, after the orbits for \tnos\ like \texttt{o3e13} were identified with multi-month arcs (\S~\ref{sec:linkage}), pointed recoveries (orange boxes) were made. Note that the pointed recoveries are not centered on \texttt{o3e13}, as the recovery pointings were chosen each lunation to encompass as many \ossos \tnos\ as possible per integration. Dots indicate observations of \texttt{o3e13} (labelled by overall lunation for clarity; blue dots: 2013, orange dots: 2014), red line with red arrows shows the position of \texttt{o3e13} from the survey start in Jan 2013 through 2015, based on the final orbit.
}
\label{fig:cadence}
\end{figure}

The \ossos project has used a dense (for outer Solar System surveys) observing cadence to provide tracking observations that enable orbital solutions within the discovery year.  In the discovery year we observed in each lunation that a given block is visible.  These observations evenly bracket the date of the block's opposition: precovery in the months before, discovery observations at opposition, recovery in the months after (Fig.~\ref{fig:cadence}).  Precovery and recovery observations on each field of each block were either a single image or a pair of images spaced by at least an hour. Each field of a block was imaged at least nineteen times in the discovery opposition.

During the discovery year the blocks were shifted over the sky at mean Kuiper belt orbital rates (Fig.~\ref{fig:cadence}). The shift rate was set at the mean motion of objects in the \cfeps L7 model \citep{Petit:2011p3938}; some 3''/hour at opposition, declining to a near-zero shift away from opposition toward the stationary point. Almost all of the sample that is present within the block at discovery is retained through the entire year by this strategy, reducing the effects of orbit shear. The shifting is done independent of any knowledge of the sky positions of the \tnos\ actually present in the field. 

The MegaCam mosaic has 13'' (70 pixels) gaps between each CCD and between the middle two CCD rows, and two larger gaps of 79'' (425 pixels) separating the first and last CCD rows from the middle two rows. To enable tracking of \tnos\ whose sky motions place them in the region overlapping these chip gaps, a dither was applied to some observations. We applied a north dither of 90'' to the observations at least once per dark run. 

A typical sequence of observations in each lunation $n$ leading up to the opposition lunation at time $t$ was thus:

\begin{itemize}
\item $t-3n$: a single observation, another north-dithered single several days later
\item $t-2n$: a single observation, another north-dithered single several days later 
\item $t-n$: a pair of observations, followed by either a single or paired north-dithered observation several days later
\item $t$: a triplet of observations, a single image a day later, followed by a north-dithered single image a day after that 
\end{itemize}
The post-opposition sequence then unfolded in reverse.  
The original cadence simulation only tested $t\pm2$, but it became possible and desirable to add $t\pm3$  during the execution of the observations (partly due to the ongoing nature of the survey operations, which could continue across CFHT semester boundaries).

The triplet of observations are the only data used for object discovery of a given block: they were acquired in the lunation that the block came to opposition. 
The triplet observations spanned at least two hours in the same night, with at least half an hour between each image of a field. 
This permits detection of sky motion by objects at distances out to $\sim\! 300$ \textsc{au}.
Due to the length of observing time required, the triplet would generally be taken on half the fields of the block one night, and on the other contiguous half on a subsequent, often adjacent, night.  The block location was shifted between these two nights, as part of our continuous shift strategy, reducing the chance that a \tno\ might be present in both half-blocks.

\subsection{2013A observations}
\label{sec:13A}

This paper covers \ossos blocks that had their discovery observations in 2013A. Forthcoming papers will cover the subsequent discovery observations (Table~\ref{tab:blocks}).
The 2013A blocks were 13AE, centered at RA $14^{h}20^{m}$, DEC $-12^{\circ}52'$ at discovery, spanning ecliptic latitude range $b=0-3^{\circ}$, and 13AO, centered at RA $15^{h}57^{m}$, DEC $-12^{\circ}30'$ at discovery, spanning ecliptic latitude range $b=6-9^{\circ}$ (Fig.~\ref{fig:13Ablocks}). Being very close to the trailing ortho-Neptune point (90 degrees behind the planet), 13AO is well placed to detect low libration amplitude 3:2 and 5:2 resonators where they are most likely to come to perihelion. The sky locations of the 13A blocks are at 44 and 30 degrees galactic latitude, comparatively close to the galactic plane for a \tno\ survey: the higher density of background stars increases the likelihood of occultations in the coming years as the \ossos objects' astrometric positions descend into the galactic plane.  

While the quality of detection is limited by the worst image in the triplet, variability in imaging conditions within blocks, and between blocks, is taken into account by the \ossos characterization process (\S~\ref{sec:characterization}). There is a single detection efficiency dependent on magnitude and moving-object motion rate for each block of observations.  
The 13AE discovery triplets were taken under some minor ($<0.04$ mag) extinction and with IQ that ranged from 0.65--0.84". 
The 13AO discovery triplets exhibited no extinction and IQ that ranged from 0.49--0.74". 
13AO exhibited a uniformly elevated sky background from low-level nebulosity, due to its proximity to the galactic plane. Although Saturn was close to the top corner of the 13AE block (Fig.~\ref{fig:13Ablocks}, blue block), the excellent rejection of off-field scattered light by MegaCam prevented much effect on the sky background of the overall mosaic, with the background of only the chip closest to Saturn affected.
All the increased sky noise is characterized by our detection efficiency ({\S~\ref{sec:characterization}).


Subsequent imaging to track the discoveries was acquired through August 2013. 
Not all discoveries were observed in every lunation due to objects falling in chip gaps or on background sources on some dates, 
faint magnitudes, or variable seeing in the recovery observations. 
In much of 2013, poor weather conditions prevented observations in sufficient IQ for us to recover the faintest objects. To compensate, from November 2013 onward we used alternative 387 s exposures in $0.8\pm0.1 ''$ seeing for single-image passes on the block. This significantly improved the ease of later arc linkage on the discoveries (\S~\ref{sec:linkage}).
Even with the occasional loss of an expected measurement, the orbital quality from the available set remains very high.

For the seven February-August lunations that the blocks were visible in 2014,
the 13AE and 13AO discoveries brighter than the characterization limit (\S~\ref{sec:characterization}) were observed with pointed recoveries;
this was possible because the high-frequency cadence in the discovery year shrank the ephemeris uncertainty
to a tiny fraction of the MegaPrime field of view.
A handful of fainter objects not immediately recovered in the first pointed recovery images were targeted with spaced triplets
of observations in subsequent lunations until recovery was successful on all of them (\S~\ref{sec:linkage}).
Generally, two observations per object per lunation were made.
The large camera field of view (FOV) allowed 2-10 \tnos\ to be observed per pointing through careful pointing choice,
ensuring that the small error ellipses of all objects avoided the mosaic's gaps between CCDs.
Each targeted pointing centre was shifted throughout the lunation at the mean motion of the discovered \tnos\ within the FOV,
ensuring the targeted \tnos\ would be imaged.
Combined with the non-linear improvement in object orbit quality (\S~\ref{sec:linkage}),
which meant not all \tnos\ required imaging every lunation,
we were able to make the necessary observations each lunation with fewer than the discovery opposition's 21 pointings.

\section{Astrometric and photometric calibration}
\label{sec:astrometry}

Systematic errors and sparsity of observation are the major limiting factor of current Solar System object astrometry.
The astrometric measurements of \tnos\ reported here are tied to a single dense and high precision catalog 
of internally generated astrometric references. Use of a high-precision catalog will minimize or eliminate
the astrometric catalogue scattering that \citet{Petit:2011p3938} encountered, allowing much more precise \tno\ orbital element determination.
This method expands on the technique of \citet{Alexandersen:2014uv}, with more images per semester.

In our \ossos calibration, each sky block
has a single coherent plate solution constructed; this is aided by the slowly
retrograding field motion, which naturally produces extensive field
overlap as the months progress, filling in all array gaps over the semester (Fig.~\ref{fig:cadence}). 
Objects with $a = 30$ \au will move eastward
$\sim\!2^{\circ}$ in a year, while sources at 60 \textsc{au}, where flux limits detection of all but the few largest objects, 
only move 0.8$^{\circ}$ per year, so the pointed recoveries in the second year of observation predominantly
overlap and enlarge the existing grid from the first year. We create
an astrometric grid with uniform photometric calibration across the
entire dataset for each block throughout our observing. We used
\texttt{MegaPipe} \citep{2008PASP..120..212G} with some enhancements.
This grid uses stellar sources that are much brighter than almost all \tnos.

The astrometry was done in three steps, resulting in three calibration levels:

\begin{itemize}
\item Level 1: individual images were
  calibrated with an external reference catalogue. This was sufficient
  for initial operations in the data pipeline, such as object
  discovery, and object recovery at the end or during each dark run.
\item Level 2: the source catalogues from the individual
  images were merged to produce a single internal astrometric catalogue,
  which was then used to re-calibrate each image. This step was repeated
  every few dark runs.
\item Level 3: the images themselves were merged to produce
  a mosaic covering an entire block. An astrometric catalogue was
  generated from this combined image and used to re-calibrate each
  individual image. This step was run at the end of each observing
  season.
\end{itemize}

The orbit classifications we provide in \S~\ref{sec:classification}, and the
information we report to the Minor Planet Center (MPC), are from
measurements relative to our final level 3 internal astrometric catalogue.

\subsection{External astrometric reference catalogues}
\label{sec:extcatalogue}

The internally generated catalogue provides a high-precision reference for our measurements; 
these highly precise measurements must then be accurately tied to an external reference system.
The 13AE and 13AO blocks were not completely within the area imaged by the Sloan Digital Sky Survey (SDSS) \citep{Ahn:2014fa}, 
which if available would have been used in preference due to its superior accuracy and depth.
Instead, 2MASS \citep{Skrutskie:2006hl} was used, with corrections based on UCAC4 \citep{Zacharias:2013cf}.  
2MASS is deeper than UCAC4 and therefore has a higher source density. 
However, there are small but significant zonal errors in 2MASS.  
When UCAC4 and 2MASS are compared, small zones of $\sim\!0.1$'' shifts between the two catalogues are apparent. 
The shifts occur with a periodicity of $6^{\circ}$ in
declination, corresponding to the observing pattern of 2MASS, which
indicates that the errors lie in 2MASS (see Fig. 2 of \citet{Gwyn:2014jg}). Therefore we use the 2MASS catalogue
which provides the source density needed to precisely link our internal catalogue to the external reference, 
corrected to the UCAC4 catalogue, which provides a more accurate translation to the International Celestial Reference System\footnote{\url{http://www.iers.org/IERS/EN/Science/ICRS/ICRS.html}}. 

\subsubsection{Proper motions}

We assessed the stellar proper motions to create our corrected astrometric catalogue.
The mean proper motion of the stars is due to the motion of the Sun
relative to the mean galaxy. Figure \ref{fig:sumgal} shows the catalogued mean proper
motion represented as vectors plotted in equatorial coordinates, 
computed by taking the median per square degree of
the proper motions of all stars in the region in the UCAC4 catalogue. 
Neighbouring vectors from each square degree are close to identical. 
\tnos\ move only a few degrees over the course of the four-year
survey, and thus differential proper motions do not measurably affect the
internal astrometry. 

\begin{figure}
\plotone{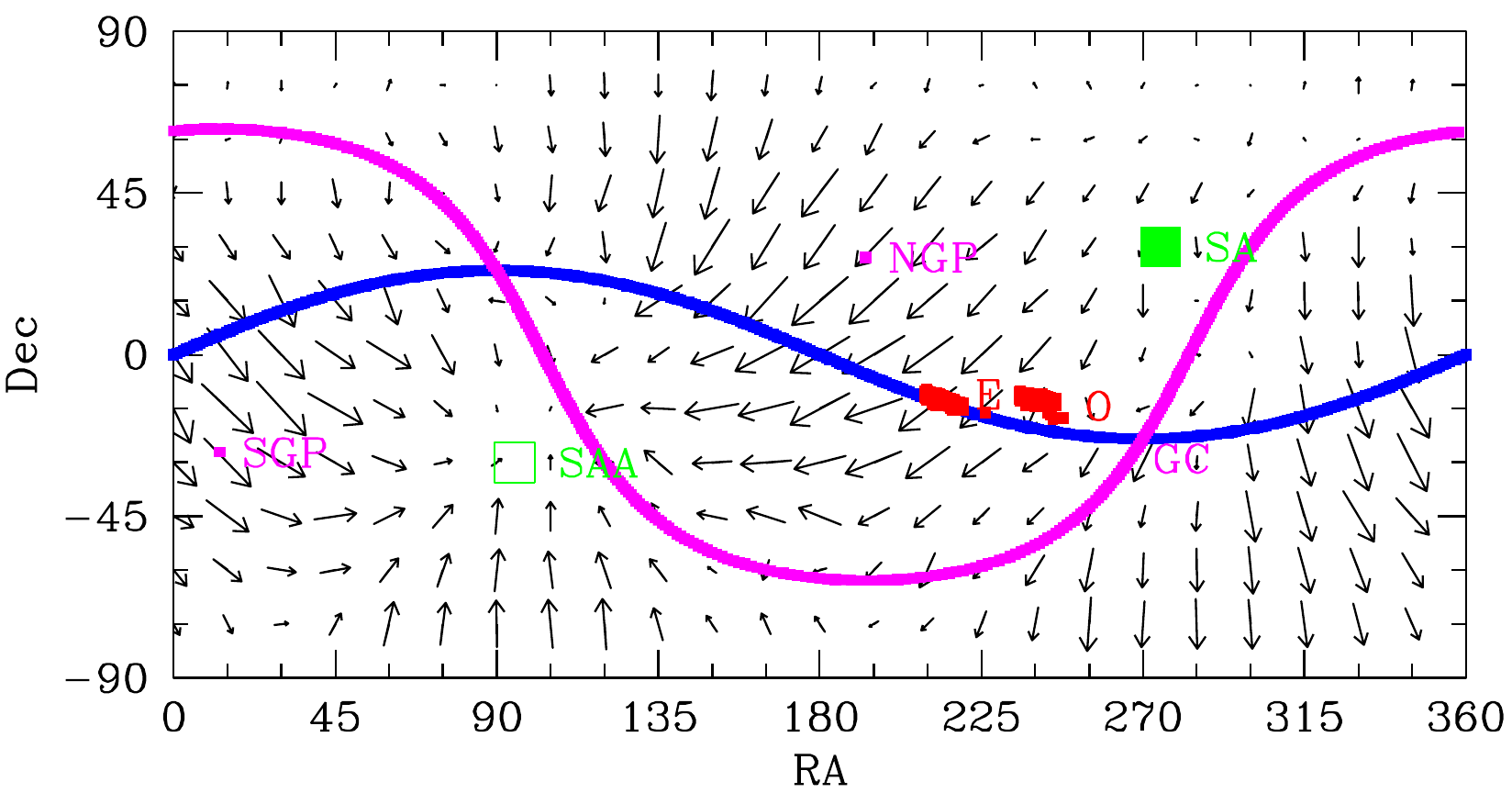}
\caption{Mean proper motion of the stars in the background astrometric catalogue on the sky. The vectors indicate the mean
  proper motion of stars: the longest vector is 40 mas/year. The Sun is moving towards the solar apex (SA) (solid green square) and
  away from the antapex (SAA) (unfilled green square). In both panels, the ecliptic is
  shown in blue, the galactic equator in magenta, with the north galactic pole (NGP), south galactic pole (SGP) and galactic center (GC) indicated. The 13AE and 13AO blocks are red.}
\label{fig:sumgal}
\end{figure}

Removal of individual stellar proper motions would improve the accuracy of the resulting
astrometric calibration. For the fainter sources that form
the majority of the UCAC4, however, the individual proper motion measurements
are too noisy. Figure \ref{fig:indivproper} shows the proper motion of
stars over a quarter of a square degree. The typical uncertainties on the
proper motions are about 10 mas, which multiplied by the 10 year
difference in epoch between UCAC4 and \textsc{ossos}, results in a 100 mas
uncertainty in position. Furthermore, the individual proper motions
are only known for the UCAC4 sources. The median annual proper motion
on the other hand (in red in Fig.~\ref{fig:indivproper}) is relatively well defined,
and could be used to apply a systematic correction between the catalogues.

The corrections were therefore applied to each image by taking a subset of the 
UCAC4 and 2MASS catalogues from Vizier \citep{Ochsenbein:2000fm},
determining the mean proper motion in that area, 
applying that to UCAC4, and matching the UCAC4 and 2MASS catalogues to each other.
Working in $0.2$ deg$^{2}$ patches, the median shift between UCAC4 and 2MASS was then
applied to 2MASS. A diagnostic plot, similar to Fig.~\ref{fig:catresid}, was produced for
each image. 
$0.2$ deg$^{2}$ provided a good compromise: at smaller scales, the number of
sources common to both catalogues drops to the point where the precision of the shift 
is less than the accuracy of the reference, leading to larger random error in the shift measurements; 
at larger scales, the zonal errors would average out, leading to larger systematic errors in the shift
measurements. The corrected result was a catalogue as deep as 2MASS, but essentially as
accurate as UCAC4.
As better astrometric catalogues become available, such as UCAC5 (Zacharias, private
communication), followed by Pan-STARRS and Gaia, it may be possible to recalibrate the data. 

\begin{figure}
\plotone{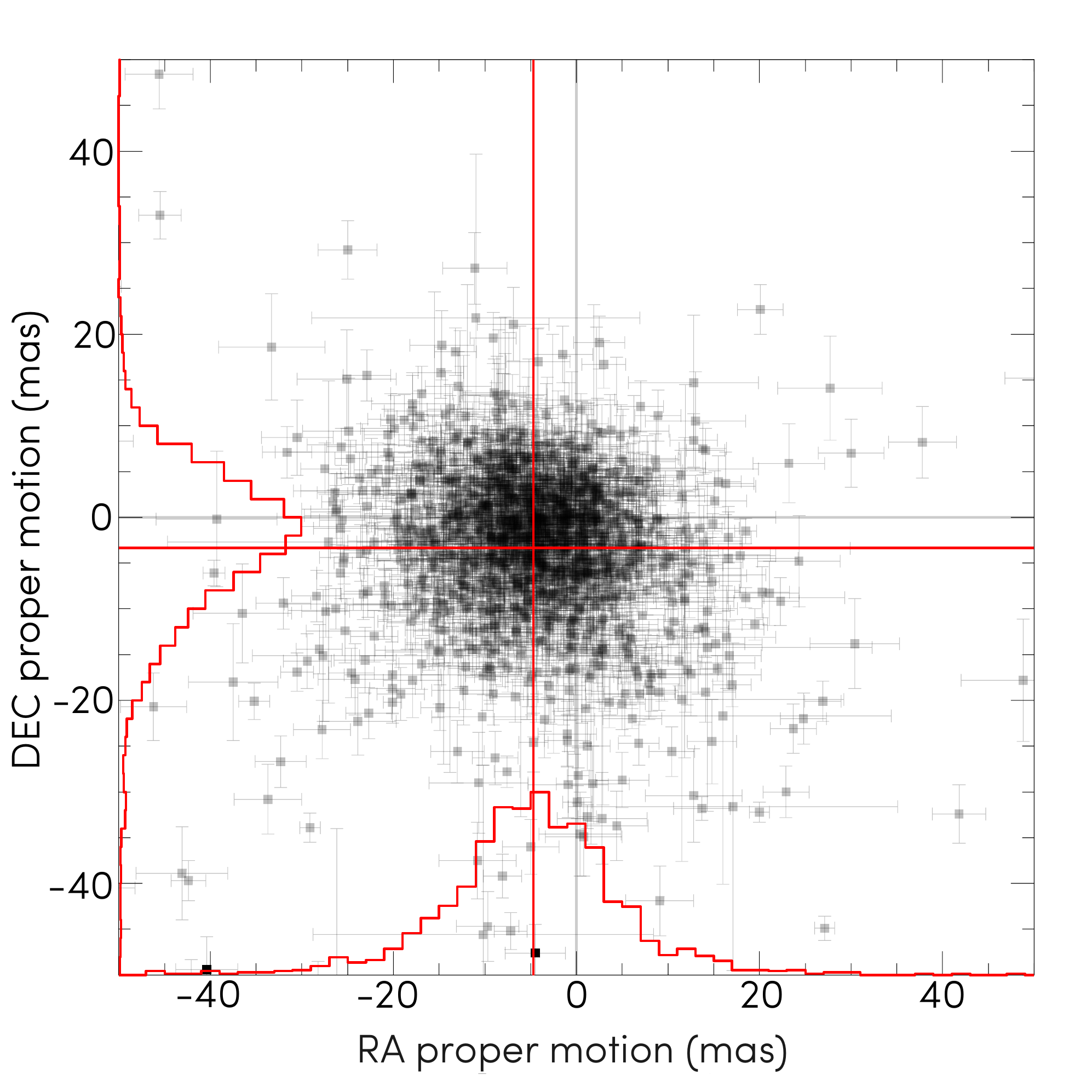}
\caption{Example of individual proper motions of stars of the background astrometric catalogue over a quarter square
  degree. The black points show the individual proper motions and
  associated uncertainties for one year as measured by UCAC4. The red
  crosshairs indicate the mean proper motion for this patch of sky: -4.7 mas in RA, -3.3 mas in DEC. 
  The red histograms shows the distribution of proper motion
  on both axes.}
\label{fig:indivproper}
\end{figure}

\begin{figure}
\plotone{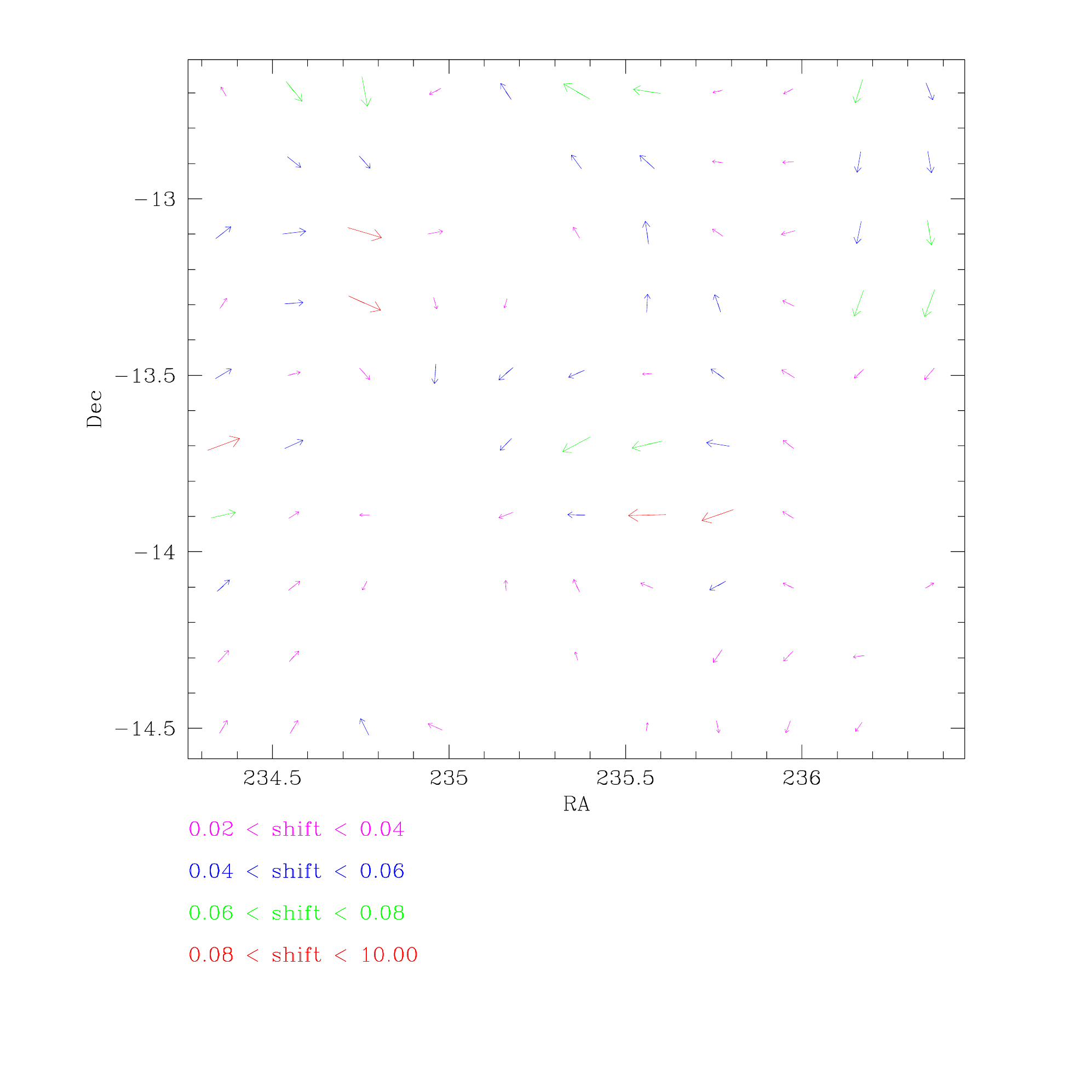}
\caption{Example of correction of the stars in 2MASS by UCAC4. The vectors
  indicate the direction and relative size of the differences between
  2MASS and UCAC4, measured in patches $0.2^{\circ}$ on a side. 
 The absence of a vector indicates that the shift was less than
0.02''. The difference in size and direction of the shifts
  between adjacent $0.2^{\circ}$ patches is small.}
\label{fig:catresid}
\end{figure}

\subsection{Level 1: individual image calibration}
\label{sec:individ}

We detrended each image as it was taken each night of the dark runs,
subtracting the bias and correcting the flat-field response. These
preprocessed images contain a basic world coordinate system (WCS) and initial zero point. 
An observed source catalogue was generated for each image with
SExtractor \citep{1996A&AS..117..393B} and cleaned of faint and extended
sources. The cleaned source catalogue was then matched to the external
astrometric reference catalogue.  Once the observed source catalogue
and the external astrometric reference catalogue were matched, the
field distortion could be measured. This process is described in detail
in \citet{2008PASP..120..212G}.  All \ossos images have at minimum
this level of calibration before any analysis is made. 
After Level 1 calibration, the astrometric residuals of the WCS are about 100
mas. 

\subsection{Level 2: merge by catalogue}
\label{sec:mergebycat}

The initial matching and fitting procedure was applied to the input
images. The computed WCS was then applied to the observed
source catalogues to convert the $x,y$ pixel coordinates to RA and
Dec. The RA/Dec catalogues were then combined to produce a merged
astrometric catalogue covering the whole block.
A given \ossos field can be observed repeatedly on a single night (\S~\ref{sec:cadence}); 
including all the images would weight some parts of each field
preferentially. Therefore, in such cases only the image with the best seeing 
was used to make the merged catalogue. 
For merging the catalogues, sources in two different catalogues were
deemed to be the same object if their positions lie within 1''
of each other, irrespective of magnitude. To avoid confusion, no source is used if it has a
neighbour within 4''. Sources often lay in more than two catalogues, 
due to the drift of pointing centers from night to night (\S~\ref{sec:cadence}); 
all matches were grouped together. 
The result was a catalogue on the original reference frame (e.g. SDSS or 2MASS, corrected to UCAC4) 
but with smaller random position errors and a higher
source density. This merged astrometric reference catalogue was then
used to re-calibrate the astrometric solution of each individual image. 
This procedure was repeated two to three times, 
until the internal astrometric residuals stopped improving.
The Level 2 calibration brought the astrometric residuals down to 60 mas.

\subsection{Level 3: merge by pixel}
\label{sec:mergebypix}

To further enhance the internally generated astrometric reference frame, 
we generated a reference catalog from stacked images.
In this step, the images with the updated Level 2 WCS in their
headers were combined using
SWarp \citep{2002ASPC..281..228B}\footnote{\url{http://www.astromatic.net/software/swarp}}, 
producing a large stacked image covering an entire survey block. SExtractor was run on this
stacked image to generate the final astrometric catalogue, and this
catalogue was used to calibrate the original images. This image stacking
step effectively combines all the available astrometric information from
each star in each image at the pixel level. In contrast, the merge by catalogue
method described in the previous section (and many other astrometric
packages) only combines information about the centroids of the astrometric sources.
The process is used to produce the final plate solution used in all
\ossos astrometry.  The internal
astrometric residuals were typically 40 mas after the Level 3
calibration, as shown in Fig.~\ref{fig:astromresid}. 

\begin{figure}
\plotone{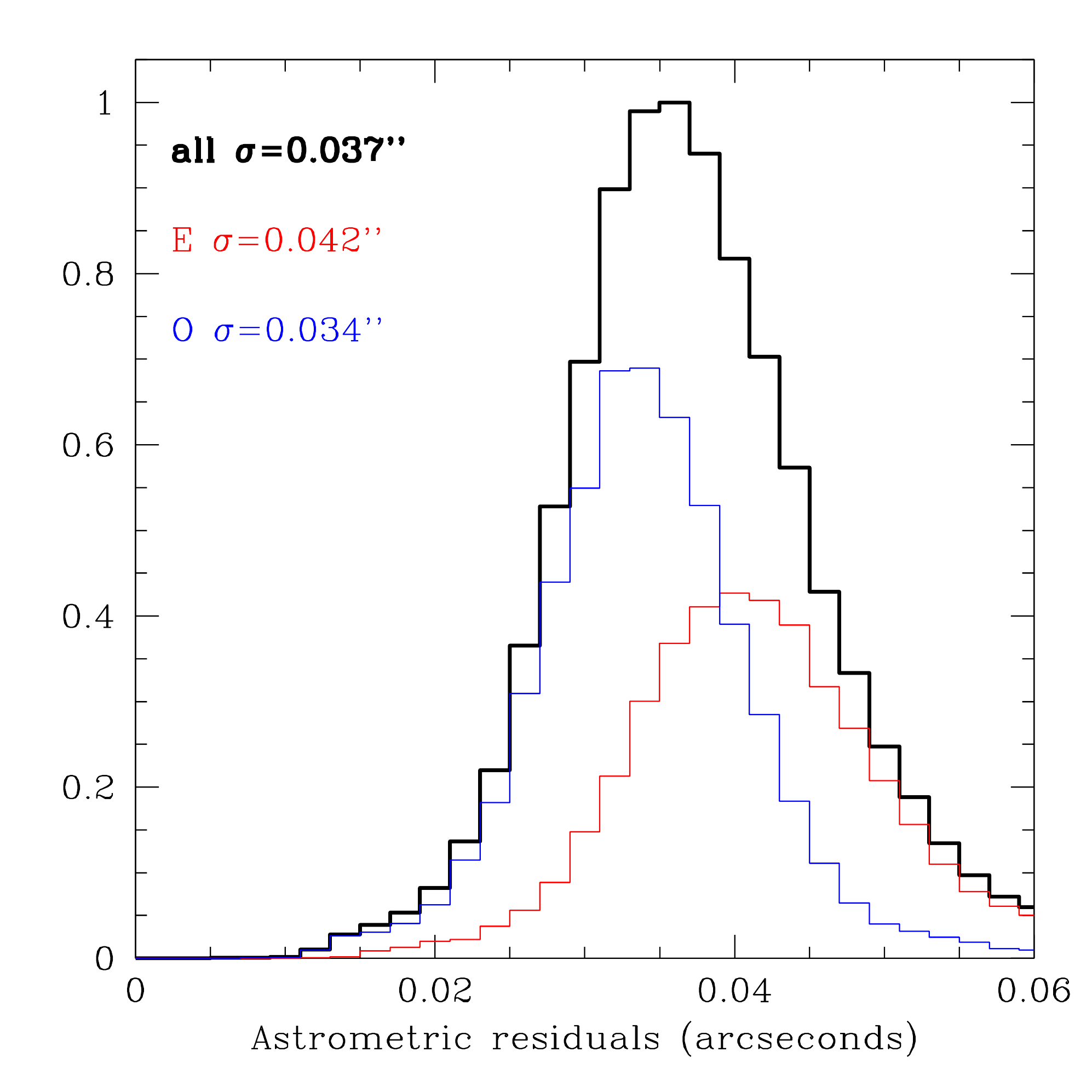}
\caption{Astrometric residuals remaining in the background astrometric catalogue for \ossos images after Level 3 (\S~\ref{sec:mergebypix}) plate solution calibration had been applied. These values are the residuals of the fixed sources in a catalogue from one image, relative to the sources in an overlapping image.
 The 13AO block is closer to the galactic plane than 13AE (\S~\ref{sec:13A}): its higher density of sources causes the small 0.008'' improvement in residuals.}
\label{fig:astromresid}
\end{figure}

However, a few nearby or high-inclination \tnos\ 
(Centaurs and some scattering \tnos) moved rapidly off
the main block. These were
re-observed in small, single-pointing patches off the main
block. These pointings are stacked separately from the main block,
resulting in a plate solution not tied directly to the
solution for the main block. These measurements are thus less precisely connected to others.  
This only occurs, however, for objects that have large intrinsic motions 
and thus have easier-to-compute orbits, 
decreasing the impact of the less precise astrometry.

\subsection{Photometry}
\label{sec:photometry}

The basis of the \ossos photometric calibration is the SDSS. 
The SDSS photometry is converted into the MegaCam system using
the following color term\footnote{\url{http://www.cadc-ccda.hia-iha.nrc-cnrc.gc.ca/en/megapipe/docs/filt.html}}:

\begin{equation}
r_{Mega} = r_{SDSS} - 0.024 (g_{SDSS} - r_{SDSS}).
\end{equation}

For typical \tno\ colors $g-r \sim\!$ 0.5--1.0, MegaCam $r$ and Sloan $r$
are thus separated by only 0.01--0.02 mags.
The MegaCam zero-point varies from chip to chip across the mosaic.
These variations are stable to better than 0.01 mags within a single
dark run and are relatively stable between dark runs. The chip to
chip variations are measured for each dark run by using any available
images which overlap the SDSS footprint; because we are measuring the
differential zero-point, it doesn't matter for this purpose if the night was
photometric.

On photometric nights, all available images overlapping the SDSS were
used to determine the overall zero-point of the camera for each
night. \ossos data taken on these nights which did not overlie the SDSS
were calibrated using a combination of the mosaic zero-point computed
nightly, and the differential chip-to-chip zero-point corrections
computed for each dark run. The nominal MegaCam $r$-band
extinction coefficient of 0.10 mags/airmass was used throughout.

The data acquired in non-photometric conditions were calibrated using overlapping images.  The
catalogues for each of the images were cross matched and the
zero-point difference for each overlapping image pair was
measured. The image overlaps are substantial; typically 2000 stars
could be used to transfer the zero-point to a neighboring
non-photometric image. The images overlapping with photometric images
were in turn used to calibrate further images iteratively until an
entire block was calibrated. At each iteration, the photometric
consistency was checked. If a pair of ostensibly photometric images
were found to have a large ($>0.02$ mag) zero-point difference, both
were flagged as non-photometric and re-calibrated in the next
iteration.

At Level 1 (\S~\ref{sec:individ}), the photometric accuracy is 0.01 mag for images on the SDSS.
For images not overlying the SDSS, the accuracy falls to 0.02--0.03 mags if the images were taken under photometric conditions.
By Level 3 (\S~\ref{sec:mergebypix}), the internal photometric zero-point calibration between images within
a block using this method is accurate to 0.002 mags RMS (Fig~\ref{fig:photomresid}). The photometric residuals with
respect to the SDSS are better than 1\% (Fig.~\ref{fig:photomresid}). Note that data are not directly calibrated with
the SDSS, but rather that the ensemble of the SDSS is used as photometric
standards.

\begin{figure}
\plotone{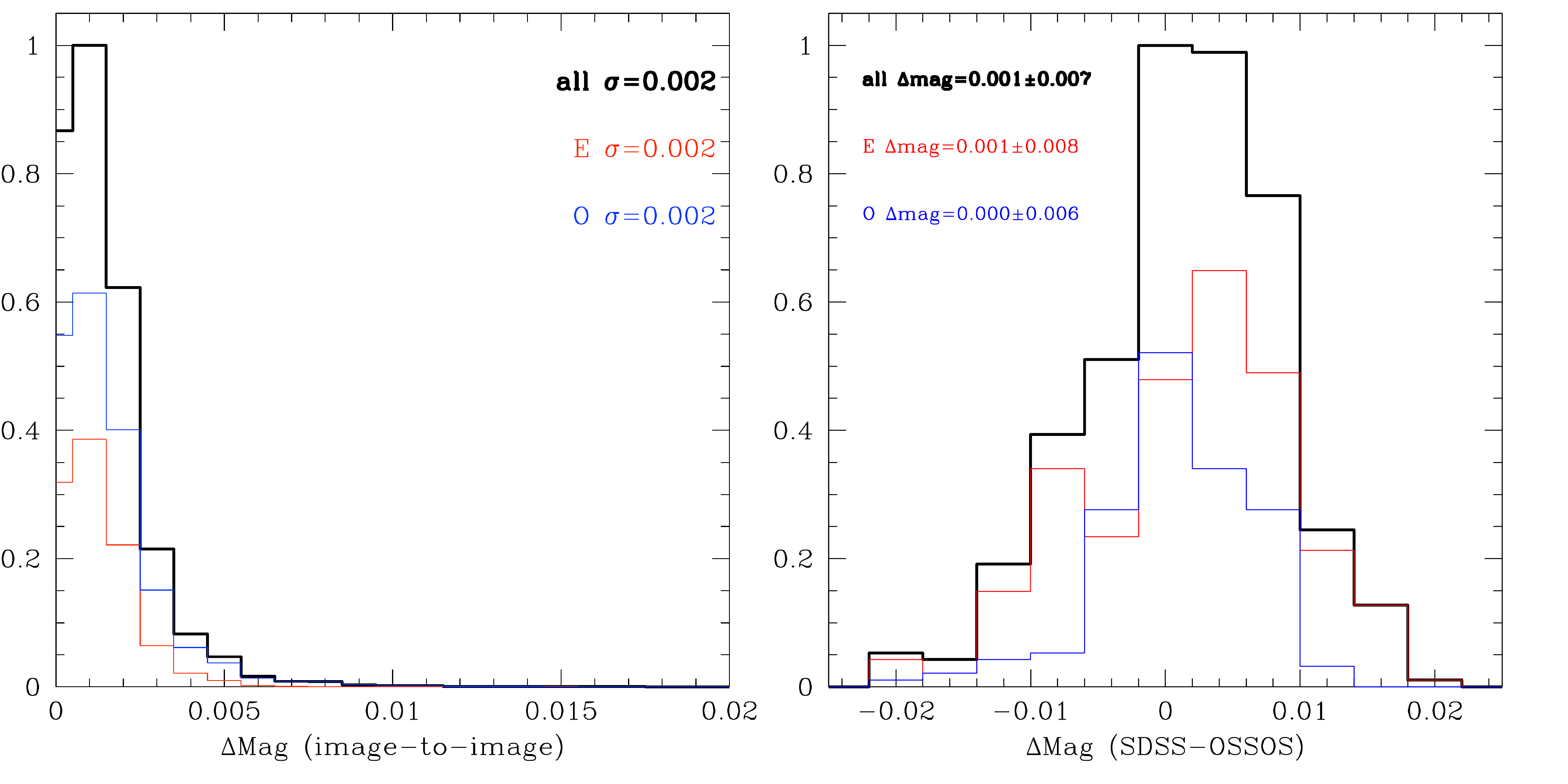}
\caption{Photometric residuals of the background astrometric catalogue of the 13AE and 13AO blocks. Left: internal image-to-image residuals; right: overall residuals with respect to the SDSS.}
\label{fig:photomresid}
\end{figure}

\section{Data processing for discovery}
\label{sec:pipeline}

The moving object discovery pipeline is designed to dig as much as possible down to the noise limit of the images, 
to find low-SNR moving targets while also generating minimal numbers of false positives. 
This strategy is critical because the steep \tno\ luminosity function
means the majority of the detections occur at low to moderate SNR.
The \ossos discovery pipeline follows the methodology described in \citet{2004MNRAS.347..471P}
and used by the \cfeps project \citep{Jones:2006jl,2009AJ....137.4917K,Petit:2011p3938}.
This uses two separate processing streams, one based on source detection using SExtractor
and the other based on identification of point-spread functions in wavelet space.
Source lists for each image in a triplet are produced, matched and stationary sources are removed, 
then the remaining sources are searched for linearly moving objects.
A few specifics of the original pipeline not described in \citet{2004MNRAS.347..471P} are detailed below.
The complete \ossos detection pipeline is open-source (\S~\ref{sec:code}). 


Matching stationary-source lists requires some choices on the criterion of a match: 
we require sources to have matching spatial alignment, similar flux, and similar size.
These constraints are scaled relative to the FWHM of the first frame in the triplet.
Additionally, when two sources in a single frame are found within one pixel of each other, they are merged.
Visual examination of the merged source lists reveals that this matching algorithm does a reasonable job
(90\% of stationary sources are matched between frames) of matching galaxy and stellar centroids.
The stationary sources are removed from further consideration.

\tno\ candidates are found in the images by trial linkages of non-stationary sources identified in the individual images.
Each pipeline searched the list of non-stationary sources it had independently compiled 
by linking sources across triplets whose position changes were consistent with
rates and angles of equatorial motion appropriate to the semester of  observation.
Apparent equatorial rates and angles of motion are dominated by the Earth's orbital motion.  
For the 13A blocks, moving objects were retained within rate cuts 0.4--15 $''\!$/hr,
at angles of equatorial motion $20^{\circ} \pm 30^{\circ}$ north of due west.
The parameters were set generously to ensure that they encompassed motions consistent with
any detectable objects within 10-200~\au of the Earth.
Because the retrograde parallactic motion dominates the sky motion, 
all orbital inclinations (0-180 deg) fall within our search space at trans-Neptunian distances: 
retrograde heliocentric orbits would be detected.

The independent output of the SExtractor-based and the the wavelet-based branches of the pipeline 
each produced their own list of candidate moving objects. 
Both methods produce large numbers of false candidates. 
However, the false candidates are mostly different \citep{2004MNRAS.347..471P};
the final moving object candidate list was therefore formed by the intersection of the two lists.
To be kept, the two lists must agree that the three sources in the candidate triplet all match in sky location to within one FWHM.
This final list was then vetted by two rounds of visual inspection.
The statistics of the entire process of automated candidate production followed by two rounds of visual inspection are given in Table~\ref{tab:pipeline}.

\section{Survey characterization}
\label{sec:characterization}

We define a trans-Neptunian object survey as {\it characterized} if
it measures and makes available its pointing history and detailed detection
efficiency as a function of apparent magnitude and rate of motion, for each pointing.
This is sufficient for luminosity function surveys \citep{Petit:2008th}. 
However, to also place constraints on the orbital distribution, a survey also needs to minimize
ephemeris bias; otherwise systematic biases can be introduced into the derived
orbital distribution \citep{Kavelaars:ws, Jones:2010bb}. 
We detail all the needed information for \textsc{ossos}. 
This provides the characterization needed by our survey simulator,
which allows quantitative comparison between proposed cosmogonic models and
the detections of the survey.

\subsection{Detection efficiency}
\label{sec:visual}

The detection efficiency of distant moving objects is a function of their apparent magnitude and their rate and direction of motion on the night of the discovery triplet observations.
We characterize this detection efficiency by implanting tens of thousands of artificial PSF-matched
moving objects in a temporally scrambled copy of the data set, and running object detection in a double-blind manner.
Additionally, we use the method of \citet{Alexandersen:2014uv} to obtain an absolute measure of the false positive rate.

First, we create a copy of the detection triple and then re-arrange the time of acquisition in the three discovery image headers,
shuffling the three images to the order $1, 3, 2$.
These images are passed through the software detection pipeline. 
Any source that is found in a time-scrambled set that was not implanted
must be false; no real outer Solar System object reverses apparent sky motion in two hours.
Any such detections thus provide an absolute calibration of the false-positive rate \citep{Alexandersen:2014uv}.
Secondly, we then plant artificial objects into this
time-scrambled copy and pass that through the pipeline.
In the 13A data, 43800 sources were implanted per block (57 per CCD) (Table~\ref{tab:pipeline}).
In the implanted copy, any detections must thus be either artificially injected or false positives; none can be real.
Characterizing the detection efficiency in the scrambled data also
avoids planted sources obscuring detection of real ones. 

Each CCD thus has three sets of moving candidates, each from running a distinct set of three images
through the detection pipeline:
\begin{itemize}
\item from the \textbf{discovery} images: potential \tnos;
\item from the temporally \textbf{scrambled} discovery images (which have no planted sources): if accepted through the next stages of evaluation these become \textit{false positives};
\item from the temporally scrambled and \textbf{planted} images: planted discoveries, which if subsequently rejected are \textit{false negatives}.
\end{itemize}

The detection pipeline produces 2268 sets of moving candidates per block 
(3 sets for each of 36 CCDs in each of the 21 fields of 13A's block grid),
which are stored in a central repository. 
The numbers of candidates detected for the planted, scrambled and potential \tno\ sets are listed in Table~\ref{tab:pipeline}.
In the 13A data, only 13--19000 of the 43800 planted candidates were recovered by the pipeline.
At the bright end, $m_{r} \sim\! 21$, the fraction of planted sources recovered by the entire process does not reach 100\%, 
as about 10\% of the sky is covered by stars at \ossos magnitudes.  
If a moving object transits any fixed source in one of the three images,
it tends not to be found by the automated search algorithms unless it is much brighter than the confusing source.
A gradual drop in efficiency occurs with increasing magnitude due to the increased frequency of
stellar/galactic crowding.
More candidates are planted with magnitudes faintward of $m_r > 23.5$, so that the eventual drop in detection efficiency is well quantified,
and in the 13A data most were planted fainter than could be detected.

The moving candidates are assessed by visual inspection, in two phases.
In the first round of visual inspection, 25287 candidates from 13AE and 18909 candidates from 13AO were assessed (Table~\ref{tab:pipeline}, ``Detected" columns).
Our interface is configured as a model-view-controller stack, using \texttt{ds9} as the windowing
GUI. The images are stored on a cloud server and image stamps retrieved as needed \citep{2013ASPC..475..243K}.
Each person is presented with the candidates from a randomly selected set; they do not know the nature of the set being inspected.
During evaluation, the set is locked to that person. A set is released back to the pool if the person exits the interface before
evaluating all sources in the set. 
Once fully evaluated, the set's metadata are updated (identifying that it was completely examined, and who inspected it) and the results of the
inspection are uploaded to the central repository. This robustly supports multiple people simultaneously working to examine
a block's discovery characterization. 
There is remarkably little variation in detection efficiency between the five people who assessed subsets of the 13A data (Fig.~\ref{fig:op-efficiency});
most importantly, there is strong agreement on the characterization threshold (specified below).

Any moving-object search approaching the noise limit will generate spurious candidates; 
the detection pipeline has proven its ability to massively reduce the number of such candidates \citep{2004MNRAS.347..471P}. 
The most common type of spurious candidate shown to people as part of the first vetting phase was due
was due to a candidate being formed from background noise popping above the noise threshold in three places, 
approximately linearly spaced with time. 
These false detections are easily recognized and rejected by visual inspection.
The second frequent spurious-candidate class were bright spots along diffraction effects that happened to align,
within the allowed angles of movement (\S~\ref{sec:pipeline}), across the image.
Table~\ref{tab:pipeline} shows how the  potential \tno\ candidates decrease from some two thousand (Table~\ref{tab:pipeline}: ``Detected: potential \tnos") to under two hundred (Table~\ref{tab:pipeline}: ``Potential \tnos\ after human review: \#1") due to this first inspection.

\begin{figure}
\plotone{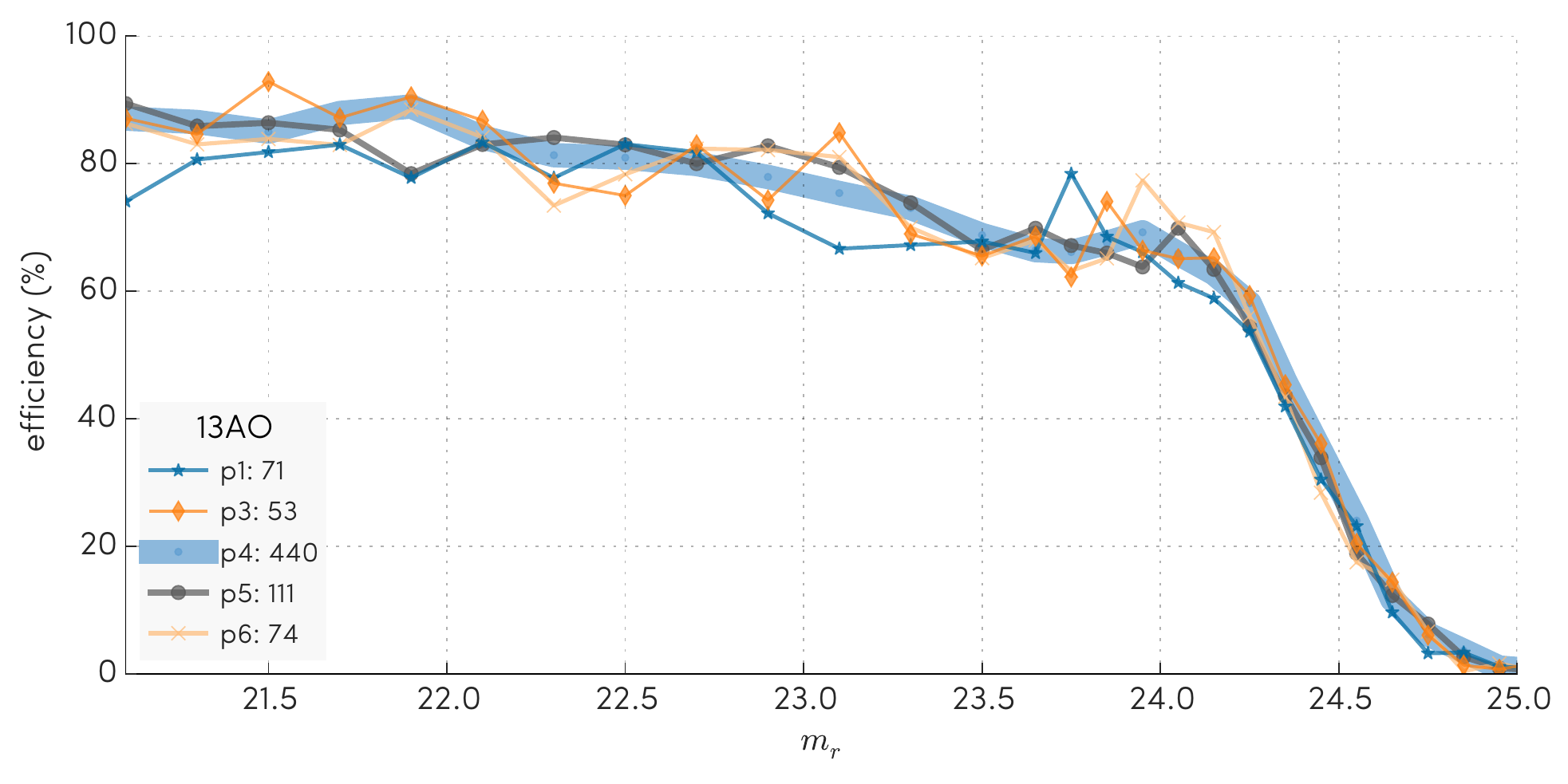}
\caption{Raw unsmoothed individual participant detection efficiencies for the first-round candidate inspection of 13AO: the fraction of artificial objects implanted in a time-scrambled copy of the discovery triplet images that are recovered by each person $p_{i}$, as a function of $m_{r}$.
The number of CCDs reviewed by each person $p1-p6$ sets the line weight for their data and is indicated in the legend. 
This shows the effect on the overall detection efficiency output from the size of the subset of 13AO that each person reviewed. 
There is agreement in detection efficiency between people, especially at the fainter magnitudes critical for characterization, where more artificial objects were planted in order to accurately characterize the roll-over and steep drop of the efficiency function.
At $m_{r} < 24.2$, the differences between people are consistent with Poisson errors.
\label{fig:op-efficiency}}
\end{figure}

The second visual inspection evaluated the remaining moving object candidates.
For resilience, this second examination was preferentially done by a different person (ensured via the metadata for each moving object candidate). 
All accepted candidates had aperture photometry measured with \texttt{daophot} \citep{1987PASP...99..191S}.
We manually assigned standard flags from the Minor Planet Center\footnote{\url{http://www.minorplanetcenter.net/iau/info/ObsNote.html}} to the photometry and astrometry of the candidates from the discovery images.
These measurements defined the discovery triplet for each object (\S~\ref{sec:appendixSS}).
The potential \tnos\ are cut by a half to a third from their previous number by this inspection (Table~\ref{tab:pipeline}: ``Potential \tnos\ after human review: \#2"). 
False positives and negatives from the whole process are given in the final two columns of Table~\ref{tab:pipeline}.
False positives are any candidates from a scrambled set that survived the second examination.
False negatives are any planted candidates that were successfully identified by the automated pipeline, 
but then (incorrectly) rejected during either of the visual inspections. 
No false positives that were brighter than the characterization limit survived the two-stage visual assessment process.
This implies that our efficiency function (discussed below) is of high accuracy and unpolluted.
The false negative rate produced during the twofold visual inspection was 0.75\%, all due to 
superposition of a candidate on a bright source or an extended background source. 
The false negatives were independent of the planted candidate magnitude and of candidate motion rate, 
instead showing a minor dependence on the sky density of extended background sources. 
This shows that about 1\% of real \tnos\ would have been rejected, and this is accounted for by the efficiency function (detailed below).



\begin{deluxetable}{c|r|rrr|rr|rr}
\tablecolumns{10}
\tablecaption{Moving-object candidates retained by the three steps of data processing for the 13A blocks of the \ossos survey \label{tab:pipeline}}
\tablehead{\colhead{Block} & \multicolumn{1}{|l}{Planted} &\multicolumn{3}{|c}{Detected by software pipeline}  & \multicolumn{2}{|c}{Potential} & \multicolumn{2}{|c}{$< m_{characterized}$} \vspace{-0.1cm} \\
\colhead{} & \multicolumn{1}{|l}{} & \multicolumn{2}{|c}{} & \multicolumn{1}{c}{} \vspace{-0.1cm} & \multicolumn{2}{|c}{\tnos\ after} & \multicolumn{2}{|c}{post-review} \\

\colhead{} & \multicolumn{1}{|l}{} & \multicolumn{2}{|c}{} & \multicolumn{1}{|c}{potential} \vspace{-0.1cm} & \multicolumn{2}{|c}{human review} & \multicolumn{1}{|c}{false}  & \multicolumn{1}{|c}{false} \\

\colhead{} & \multicolumn{1}{|l}{} & \multicolumn{1}{|c}{planted} & \multicolumn{1}{|l}{scrambled} & \multicolumn{1}{|c}{\tnos} \vspace{-0.1cm} & \multicolumn{1}{|r}{\#1} & \multicolumn{1}{|r}{\#2} & \multicolumn{1}{|l}{positives} & \multicolumn{1}{|l}{negatives}
}
\startdata
13AE	& 43800 & 13639	& 2773	& 2497  	& 119	& 54	& 0	& 133	\\
13AO	& 43800 & 19957	& 3292	& 2038	& 154	& 50	& 0	& 149	\\
\enddata
\tablecomments{
As detailed in \S~\ref{sec:visual}, the data processing pipeline (\S~\ref{sec:pipeline}) finds three sets of candidates: PSF-matched \textit{planted} objects, \textit{scrambled} candidates due purely to chance alignment of non-Solar System sources, and the set of \textit{potential} \tnos\ (from the unaltered discovery triplet). Two rounds of visual review (detailed in \S~\ref{sec:visual}) reject many of the assembled candidates. Potential \tno\ candidates retained through both rounds, listed under ``\#2", are our discoveries (Tables~\ref{tab:discoveries} and \ref{tab:unchar_discoveries}). Candidates retained after visual review that are from the detected \textit{scrambled} set are \textit{false positives}: none were brighter than the characterization limits (Table~\ref{tab:characterisation}) for their block, implying the detection efficiency function (Eq.~\ref{eq:etasquare}) is highly accurate. About 0.75\% of the detected planted candidates were rejected during visual inspection; these \textit{false negatives} were due to one or more points of the candidate falling coincident with a background source.
}
\end{deluxetable} 

At a certain magnitude depth in the images, about $m_{r} \sim\! 24$ for \textsc{ossos}, the SNR and thus the
efficiency with which we can detect sources rapidly falls off, setting a natural completeness limit in magnitude. \citet{2004MNRAS.347..471P} determined
that fainter than $\sim\!40$\% efficiency,
a person is no longer confident that the pipeline's moving candidates are real; a small error in the characterization at these low efficiencies would result in a large effect in the subsequent modelling.
After all the candidate sets for a given block were examined (Fig~\ref{fig:op-efficiency}), a function
was fitted to the aggregate of the raw efficiencies produced from each person blinking the planted sets
of the 756 chips per survey block (Fig.~\ref{fig:efficiency}). 
The crucial efficiency versus magnitude behavior was fit to the formulation (shown graphically in Fig~\ref{fig:efficiency})
\begin{equation}
\label{eq:etasquare}
\eta(m_r) = \frac{\eta_o - c (m_r - 21)^2}{1 + \exp{\left(\frac{m_r-m_L}{w}\right)}}
\end{equation}
where $\eta_o$ is roughly\footnote{$\eta_o$ is the efficiency at $m_{r} =21$ in the case where $\exp((21-m_{L})/w) <<< 1$.} 
the efficiency at $m_{r} =21$, $c \sim\! 0.5\%$. Eqn.~\ref{eq:etasquare} quantifies the
strength of a quadratic drop, which changes to an exponential falloff over a
width $w$ near the magnitude limit $m_L$, similar to that used by \citet[eq. 2]{Gladman:2009cx}.
This function better fits the \ossos detection efficiency than does the frequently used hyperbolic tangent function \citep{Gladman:1998dy, Trujillo:2001hg}.
The parameters we obtained for the motion-rate range 0.5--7$''\!$/hr for 13AE were 
$\eta_{o} = 0.89$, $c = 0.027$, $m_{L} = 24.17$, $w = 0.15$, 
and for 13AO were 
$\eta_{o} = 0.85$, $c = 0.020$, $m_{L} = 24.62$, $w = 0.11$.

We used this fit to set our \textit{characterization limit}: the magnitude above which we have both high
confidence in our evaluation of the detection efficiency, and find and track all brighter objects.
This is not at a fixed-percentage detection efficiency, 
unlike some previous surveys \citep{Jones:2006jl, 2009AJ....137.4917K, Petit:2011p3938, Alexandersen:2014uv}, 
but rather set more stringently at the apparent magnitude where 
\ossos ceased reaching 100\% tracking efficiency due to low flux.
In practice this was usually close to the magnitude where the detection
efficiency falls to 40\% (see Fig.~\ref{fig:efficiency}). 
The characterization limit is dependent on the moving object rate of motion: our limits are listed in Table~\ref{tab:characterisation}.

\begin{deluxetable}{ccc}
\tablecolumns{3}
\tablecaption{Characterization limits for the 13A blocks of the \ossos survey \label{tab:characterisation}}
\tablehead{\colhead{Motion rate ($''\!$/hr)} & \colhead{Characterization limit ($m_{r}$)} & \colhead{Efficiency at limit (\%)} }
\startdata
\cutinhead{13AE}
	0.5--8.0	&	24.09	&	37	\\
	8.0--11.0	&	23.88	&	40	\\
	11.0--15.0	&	23.76	&	41	\\	
\cutinhead{13AO}
	0.5--7.0	&	24.40	&	55	\\
	7.0--10.0	&	24.33	&	41	\\
	10.0--15.0	&	24.17	&	41	\\	
\enddata
\end{deluxetable} 

Fig.~\ref{fig:efficiency} illustrates the variation in sensitivity to different angular rates of sky motion.
Our survey is optimized for detection of objects at Kuiper belt distances: 
this is reflected in the greatest detection efficiency for objects when they are moving with rates of 0.5--8 $''\!$/hr.
This gives \ossos sensitivity to distances out to $\sim\!300$ \textsc{au}, at which, on a circular orbit, an object would move $\sim\! 0.5$ $''\!$/hr.
Sensitivity to close, fast-moving objects ($>10$ $''\!$/hr) is comparable through $m_{r} = 23.5$; 
it drops to 40\% detectability a few tenths of magnitude brighter than for objects at Kuiper belt distances (Fig.~\ref{fig:efficiency}).
As an additional proof of the survey's sensitivity to Centaurs, the proximity of Saturn to the 13AE block
placed a few known satellites on one field of 13AE. 
Our analysis recovered the irregular satellite Ijiraq at 9.8~\au (Fig.~\ref{fig:13Ablocks}), 
the only moon above the 13AE magnitude limit, exhibiting some minor and expected elongation along its direction of motion.

All objects listed in the MPC that fell on the survey coverage of the discovery triplets were recovered,
as seen by the overlapping of symbols in Fig.~\ref{fig:13Ablocks} and noted in Table~\ref{tab:discoveries}.
While 2003 HD57 was very close to the survey coverage (Fig.~\ref{fig:13Ablocks}), it was not within the discovery observations: 
this object fell two pixels south of the first image of the 13AE discovery triplet. 
These recoveries of known objects aid our confidence in our measured detection efficiency. 

\begin{figure}
\plotone{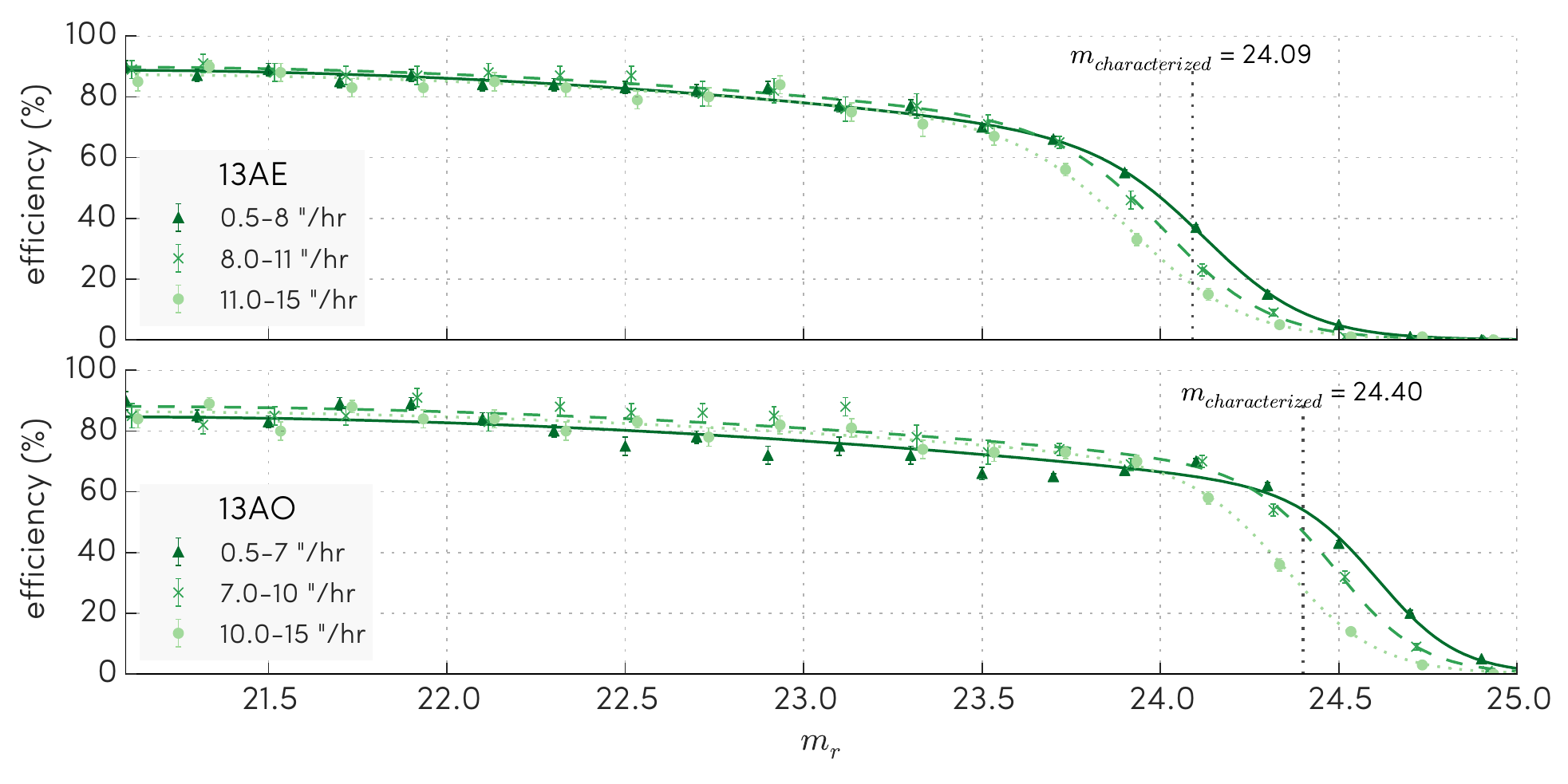}
\caption{Total combined \ossos detection efficiency in each 13A block: fraction of planted sources recovered by the overall data reduction as a function of magnitude and rate of apparent sky motion.
The efficiency begins below 100\% due to loss of sources to merges with background sky sources and to chip gaps.
Background confusion gradually increases for fainter magnitudes.
Faster-moving objects are more affected by movement off the field during the temporal span of the discovery
triplet. 13AO had better IQ during the observation of the discovery triplet,
pushing its limiting magnitude deeper.
\label{fig:efficiency}}
\end{figure}

\subsection{Survey simulator}
\label{sec:simulator}

To be usefully compared to the observed orbital distribution, a model of the \tno\ orbit distribution must be biased in the same way as the observed sample. 
Although free of ephemeris bias, the \ossos pointing history (\S~\ref{tab:pointings}) and flux limits create a biased view of the intrinsic population.
These biases are precisely modelled by the \ossos survey simulator.  
Our approach is primarily one of model {\it rejection} rather than fitting. 
The simulator selects a set of detected objects out of a given orbit model. 
This survey-biased sample of the model orbital distribution forms a valid statistical comparison to the \ossos \tno\ discoveries.
The decision about how to compare the simulated set of detections to 
the \ossos set of \textit{characterized} discoveries, those brighter than their block's characterization limit,
is then a statistical problem. 
Various approaches are described in \citet{2009AJ....137.4917K, Petit:2011p3938, Gladman:2012ed, Alexandersen:2014uv, Nesvorny:2015ik}.

The simulator is similar to that described in \citet{2009AJ....137.4917K} and \citet{Petit:2011p3938}.
An orbit distribution model is exposed to the survey biases via the survey simulator.
Each model object is specified by a set of orbital elements and an absolute magnitude in some reference passband.  
An improvement in the \ossos survey simulator is that each model object is also assigned a surface reflectance, 
specifying that model object's color in all filters. 
Further detail on model object apparent magnitudes is given in \S~\ref{sec:appendixSS}. 
The current implementation of our simulator can take into account rotational variability; 
we currently have insufficient information in the \ossos discovery and tracking data to take advantage of this improvement. 
Our discovery observations, covering a two-hour baseline, do at least potentially measure variability over a moderate fraction 
of typical \tno\ rotation periods of 4--14 hours \citep{2009A&A...505.1283D,2013AJ....145..124B}.
A follow-up program to comprehensively measure rotational variability for the \ossos discoveries  
would allow us to assign a light curve to model objects and use this capability of the simulator. 
The survey simulator can also apply the survey biases of other characterized surveys to the input orbit model, 
if the discovery and tracking circumstances of the additional surveys are available.


Determining the intrinsic size of a \tno\ sub-population is an important model constraint. 
Once a model distribution has been chosen, the simulator can be used to create a 
model-dependent estimate of the size of the intrinsic population.
The simulator will provide as many detected model objects as desired.
When the same number of model detections as were found by the input survey is achieved,  
the number of model objects that were checked is an estimate of the intrinsic \tno\ population.

\section{Orbits}

The loss of discovered objects due to ephemeris bias results in a biased view of the orbital distribution \citep{Kavelaars:ws, Jones:2010bb}.  
The \ossos goal is to eliminate this bias by tracking virtually all outer Solar System detections with
magnitudes above each block's characterization limit.
This was achieved for all objects above the 13A characterization limits.

\subsection{Recovery success and orbit quality}
\label{sec:linkage}

Objects found by \ossos must have their many observations converted into an orbit.
Following discovery in the opposition triplet, we knit together observations of each \tno\ from every
lunation into longer orbital arcs, starting within the discovery lunation and working outward in time.
This iterative procedure started by sending the discovery arc 
to the archival search tool \texttt{Solar System Object Image Search}
\citep{Gwyn:2012gv}\footnote{\url{http://www.cadc-ccda.hia-iha.nrc-cnrc.gc.ca/en/ssois/}},
to query for further available \ossos imaging containing the \tno. 
This tool identifies all available archived imaging, but as \ossos is deeper than most previous
wide-field imaging work, we have not yet made use of other datasets.
Starting the initial search by only querying for observations near in time to the discovery epoch kept on-sky uncertainties below 30$''$, 
minimizing the number of images to examine \citep[Fig. 1,][]{Jones:2010bb}.
We then visually identified the \tno\ within or near the predicted 1-$\sigma$ on-sky error ellipse
by comparison with \ossos images of the same piece of sky at a different time.
The \ossos observing strategy of slowly moving pointings (\S~\ref{sec:cadence}) yielded large numbers of these comparison images.
The resulting astrometry was then fed back into the search tool  to request 
more \ossos imaging in dark runs further from the detection triplet. 
We iterated until an arc over the entire discovery year was assembled.
Extending each 13A \textsc{ossos} object's arc with all the images taken in the 13A discovery semester,
an arc of 150--183 days, yielded preliminary orbits with fractional
semi-major axis uncertainty of $\sigma_{a} \sim\! 0.1-1$\% (Fig~\ref{fig:daovera}).
The small orbit uncertainty was produced by the combination of long arcs in the discovery opposition, 
frequent sampling, and the high-precision astrometric solution (\S~\ref{sec:astrometry}). 
This is an order of magnitude better than that obtained by \citet{Petit:2011p3938}.

Even though the locations of the objects were unknown when the first-semester observation suite was acquired,
the slow drifting of the blocks at Kuiper belt mean-motion rates retained almost all objects within the observations.
Independent of its characterization limit (\S~\ref{sec:visual}), each block has a {\it tracking fraction}:
what fraction of the objects above the characterization limit were recovered outside
of their discovery triplet and generated a high-quality orbit.
We recovered 100\% of our discoveries that were above the characterization limit in both 13A blocks.

The second year of \ossos observations provided astrometry that would allow classification of the orbit (\S~\ref{sec:classification}).
The first-year orbits provided such accurate ephemeris predictions 
(sub-arcminute 1-$\sigma$ on-sky error ellipses: predominantly $<10''$)
 that recovery was almost always immediate in the observations from the first lunation of the second opposition.
Those few objects which sheared off the block during the discovery year still had observational arcs spanning at least several lunations.
In these cases the uncertainty at the start of the observations the following opposition were $\sim\!30'$, 
and a manual, visual search resulted in the recovery of the object (\S~\ref{sec:13A}).
Initial recovery of the 13A discoveries in 2014 extended their arcs to $\sim\!360$ days, dropping the fractional uncertainty
in semi-major axis by a factor of 2-3 (depending on which lunations the objects were seen in 2013)
to $\sigma_{a} = 0.03\%-0.3\%$ (Fig.~\ref{fig:daovera}).
Later extension of the arc through 2014 brought the 13AE objects to a median $\sigma_{a} = 0.03$\%
and a median $\sigma_{a} = 0.07$\% for 13AO;
the difference is due to the existence of more observations per dark run for the 13AE block.
Some objects in particular converged quickly to $\sigma_{a} < 0.1$\%; by early in 2014A,
nearly half the objects in 13AE, particularly cold classicals, reached sufficiently high orbit quality (Fig~\ref{fig:daovera})
that only sparse sampling throughout the remainder of the semester was required (\S~\ref{sec:13A}).
The total number of observations on the objects varied between 14 and 55, though the median was 26; 
the number of observations is somewhat correlated with orbital quality (Fig~\ref{fig:daovera}), 
but the distribution of those observations in time is also important for the convergence of $\sigma_{a}$.
These two year observing arcs are nominally sufficient to create our final orbit estimate.

We note that the figure-of-merit $\sigma_{a}$ is only a useful approximation that does not
capture all aspects of orbit quality. 
For example, resonant libration amplitudes \citep[discussed for \ossos in][]{Volk:2015} have uncertainties that while dominated by 
$\sigma_{a}$, also depend on $e$ and the accuracy of angles like the ascending node $\Omega$ and the pericenter's longitude $\varpi$.
Location within the resonance also matters: an object with orbital elements on the edge of a resonance
might need a much smaller $\sigma_{a}$ to determine
the libration amplitude to $10^{\circ}$ precision than if its elements were near the center of the resonance.


\begin{figure}
\plotone{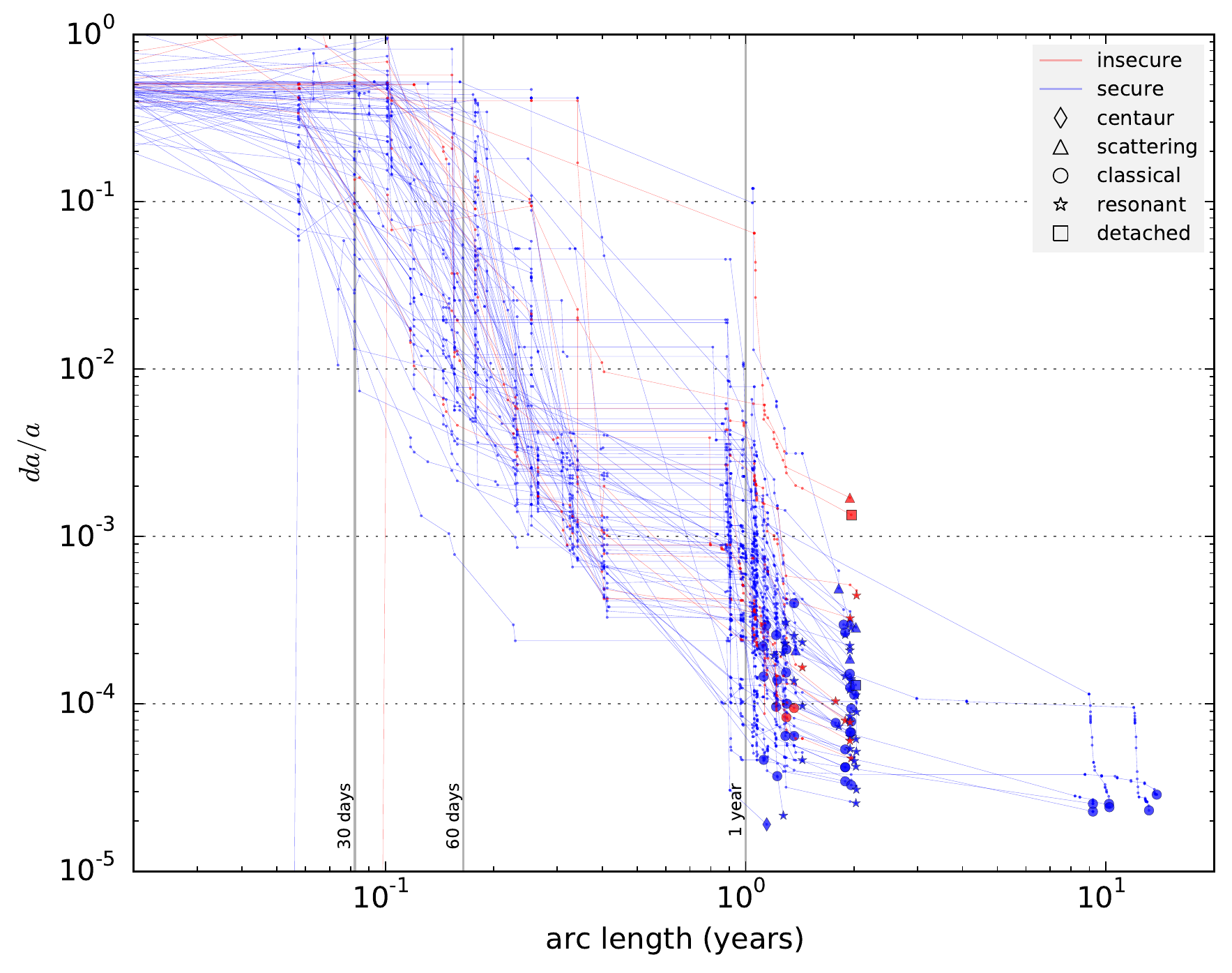}
\caption{Fractional semi-major axis uncertainty $\sigma_{a}$ of \ossos objects as a function of arc length, 
as approximated using the \citet{Bernstein:2000p444} algorithm, for each astrometric measurement made by \textsc{ossos}. 
Final orbit classifications (end symbol on each object's line) are from $10^{7}$ year integrations (\S~\ref{sec:classification});
a classification is found to be secure (line color) when the integrations of its extremal orbit-fit 
solutions and of its best-fit orbit solution all receive the same classification.
These are also listed in Table~\ref{tab:discoveries}. 
Previously discovered objects with decade-long arcs cluster at lower right.
\label{fig:daovera}}
\end{figure}

Our subarcsecond astrometry on moving targets travelling several degrees across the sky 
is a major factor in the high quality of the \ossos orbits.
It is substantially due to the use of a single astrometric solution
over the entire area that a given block traces out over the two years of the survey (\S~\ref{sec:astrometry}).
The high quality of the \ossos astrometric catalogues eliminates nearly all of the astrometric
catalogue scattering that \citet{Petit:2011p3938} encountered: the median \ossos astrometric residuals
around the best orbit fit are twofold lower than \citet{Petit:2011p3938}'s typical orbit-fit residuals of 0.25'' (Fig.~\ref{fig:best_fit_resid}). The catalogue approaches what the future Gaia catalogue will provide in absolute astrometry.
Only for our very brightest objects is the astrometric scatter in the solution slightly worse than the
centroid uncertainty --- at the characterization limit, the residuals are centroid-limited.
Further improving the internal astrometric solution's scatter will therefore not result in improvement
to the \ossos orbit precision.

\begin{figure}
\plotone{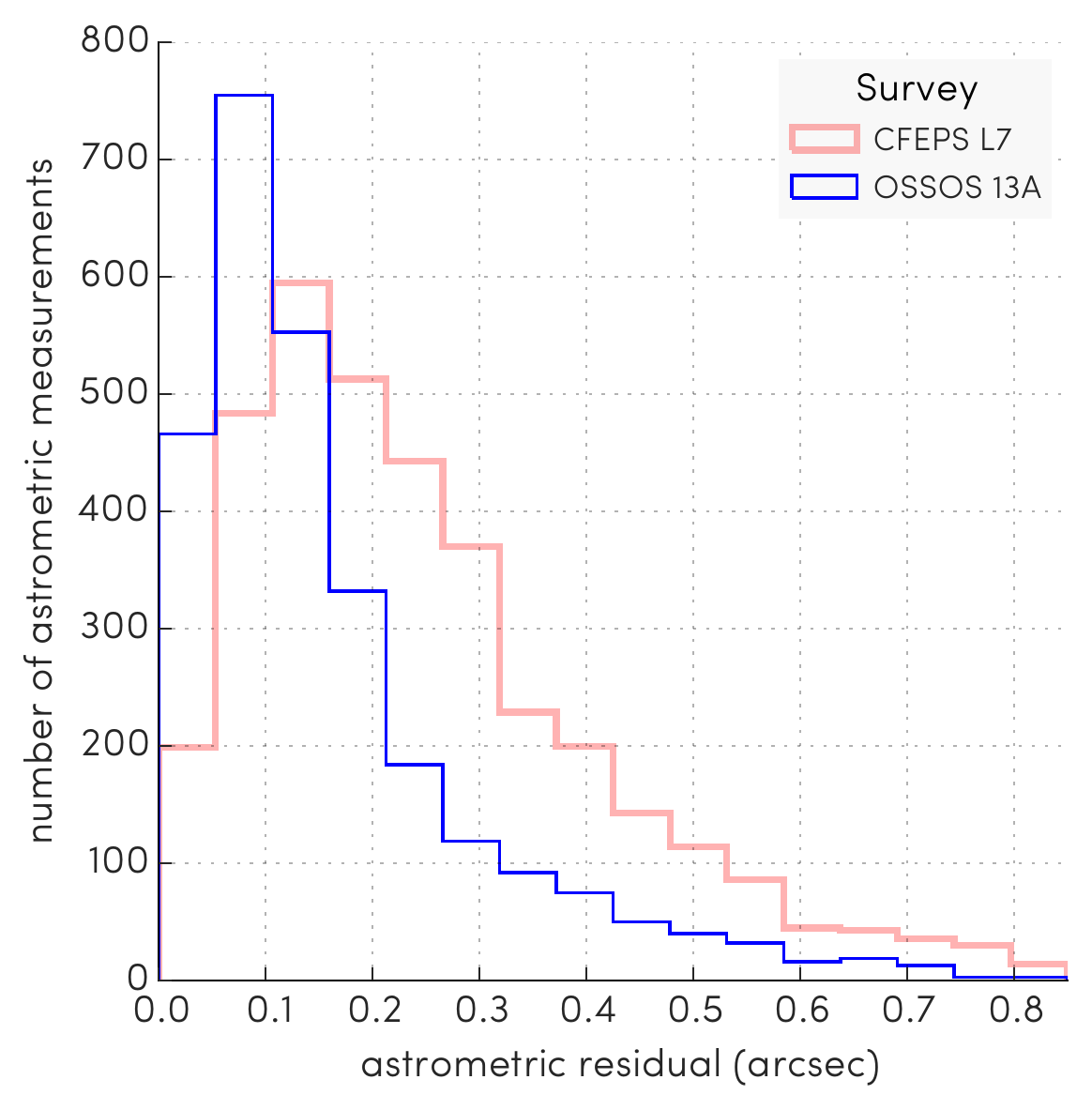}
\caption{The astrometric scatter of \ossos observations (median 124 mas; 2872 measurements) relative to the
\citet{Bernstein:2000p444} algorithm best-fit orbits for the 13A discoveries. 
For reference, the distribution from \citet{Petit:2011p3938}'s detections (median 216 mas; 3643 measurements) is shown.
Note that most of the detections have SNR $<10$, and so the measurement accuracy is essentially
the centroiding scatter on the faint targets:
the \ossos plate solutions are so accurate that catalog scatter has become irrelevant.
\label{fig:best_fit_resid}}
\end{figure}

\subsection{Orbit classification}
\label{sec:classification}

The classification scheme for the \ossos detections is that described by \citet{Gladman2008},
which we briefly summarize here.
A best-fit orbit for each \ossos detection is computed using the \citet{Bernstein:2000p444} algorithm.
Maximum and minimum semi-major axis orbits consistent with the observations are found by searching
the parameter space, starting at the best fit,
via a Monte-Carlo method to identify an orbit in the 6-D parameter space
with the two extremal values in $a$,
which have residuals no worse than 1.5 times the best-fit orbit's residuals.

These three barycentric orbits are converted to heliocentric, ecliptic coordinates and integrated forward in time for $10^7$ years using the \texttt{rmvs3} subroutine
within the SWIFT integrator package \citep{Levison1994}; the planets' positions are taken from the JPL Horizon's service \citep{jplhorizons} for the epoch of the orbit fit.
These integrations are first checked for resonant
behaviour, defined as libration of a resonance angle of any resonance up to $30^{th}$ order within 2\% of the
object's average semi-major axis (see further discussion in \citet{Volk:2015}). 
Objects with $a<30$ \au and not resonant with any planet are classified as Centaurs.
An object is classified as scattering if its semi-major axis varies by more than 1.5 \au during the integration.
Objects with constant semi-major axis over the $10^7$ year period are classified as detached if they have
$e>0.24$ or as classical if they have $e<0.24$. 
As in \citet{Gladman2008}, this eccentricity division is arbitrary to maintain a distinction between 
objects with pericentres decoupled from Neptune and non-resonant low-$e$ \tnos.
Classifications are considered \textit{secure} if all three integrations for an object receive the same classification.
The fraction of securely classified objects that we achieve within two years is 94\%.
In contrast, objects in the Minor Planet Center ensemble that have been observed since discovery with sparser cadences 
lack classifiability within this timeframe \citep{Gladman2008}.

However, orbital insecurity is still a property of some characterized, fully tracked \ossos discoveries.
This is not due to poor-precision measurements;
even with excellent ground-based data in $0.5''$ seeing, there is a
fundamental degeneracy to a suite of orbits, all of which produce the
same short-arc behavior.
Most \ossos objects were not secure in their first year of observation,
when orbital arcs were usually 4 or 5 months long.
The addition of even a single dark run in the second year usually resulted in
a classical-belt object identification being secure.
Secure resonant identification usually required the full suite of dark
runs in both observation years.
During the four-year duration of \textsc{ossos}, insecure objects will continue
to be tracked until their classifications become secure; for example,
ten of the 13A discoveries received another measurement in January and in March 2015
to improve orbital quality. 
All of the currently insecure classifications are due to proximity to resonances
of at least second order.

\section{Discoveries}
\label{sec:discoveries}

Fig~\ref{fig:orbdiscoveries} shows that the general pattern of the orbital elements of the 85 first-quarter
\ossos detections are consistent with the known populations of the Kuiper belt.
The majority of the objects are detected at heliocentric distances $d$ (top panel) from 28 \textsc{au}, 
the perihelia $q$ of the lowest-$q$ resonant \tnos, smoothly out to 45 \textsc{au}. 
In the $d$=28-45 range the inclination distribution is
that of the dynamically hot objects; the few low-$i$ objects are
the tail of the Gaussian distribution of the dynamically hot objects, down towards $i$=0.  
At $d\sim\!40$~\au there is the sudden appearance of the dynamically
cold classical belt, which we discuss further in \S~\ref{sec:classical}.
The relative importance of the classical belt is muted in our sample 
due to the relatively shallow depth of the ecliptic 13AE block.
Only four of our detections have $d>50$~\textsc{au}.

In $a/i$ and $a/e$ space we see the usual spread to large orbital
inclinations, predominantly detected in our moderate-latitude 13AO block,
and the tail of large-$e$ orbits that correspond to members 
of the scattering, resonant, and detached populations detected near 
perihelion. 
The implications of these detections are discussed for the scattering population in \citet{Shankman:2016hi}, 
and for the resonant populations in \citet{Volk:2015}.
Two resonant objects are the lowest inclination yet found in their resonances: 
\texttt{o3e19} (2013 GR$_{136}$) at $i = 1.6^{\circ}$ in the 7:4, and 
\texttt{o3e55} (2013 GX$_{136}$) at $i = 1.1^{\circ}$ in the 2:1. 
In contrast, no such low-$i$ objects have yet been detected in the 3:2 Plutinos.

Some of the \tnos\ in the \ossos discovery sample were previously discovered in other surveys: 
seven 13AE and one 13AO object link either to one-night observations from the \cfeps survey,
or to objects of varying arc length in the \href{www.minorplanetcenter.net/iau/MPCORB.html}{public catalogue} at the Minor Planet Center, 
providing arcs to objects first observed 9 to 13 years ago.
Their listings in Table~\ref{tab:discoveries} have an \ossos ``PD'' suffix. 
Their MPC designations are for discovery years significantly earlier than 2013 -- however,
they now benefit from having a well-characterized detection study.
For five of the previously-observed objects, the astrometric quality of the earlier observations were lower than what we report here. 
We note that if we back-predict the position of these objects, using only our astrometry, to 8--10 years ago, 
we are within 10 arcsec of the previous measurements.
Incorporating these earlier observations improved the $\sigma_{a}$ of these objects by a factor of only about 2-3 over those of the best 17-month \ossos orbits. 
The importance of the survey strategy's emphasis on tracking all objects (\S~\ref{sec:cadence}) is shown by how it allows us to re-find untracked objects from previous surveys that are on the wrong orbits. For example, for 2002 GG$_{166}$ (\texttt{o3e01}), adding our well-sampled arcs extensively modified the orbit from the initial lunation-long arc. 2002 GG$_{166}$ was initially published\footnote{MPEC 2002-L21: \url{http://www.minorplanetcenter.net/mpec/K02/K02L21.html}} as a Plutino: here it becomes a Uranus-crossing scattering object (Table~\ref{tab:discoveries}). 

Use of \tno\ orbits as statistical constraints on models of the formation and evolution of the Solar System is dependent on being certain of the detection characterization of those objects. For \textsc{ossos}, the objects whose flux at discovery is fainter than the characterization limit are not included in our model analysis; they are listed in Table~\ref{tab:unchar_discoveries} and have been reported to the MPC. 
Over the full $\sim\!170$ deg$^{2}$ survey we anticipate detecting $\sim\!500$ outer Solar System objects brightward of our characterization limits. The current rate of detection of \tnos\ in the \ossos survey is roughly consistent with expectations given our achieved characterization limits, ecliptic latitude locations surveyed and the currently known luminosity function of \tnos\ \citep{Fraser:2014vt}.  The 13AE discovery rate (49 objects in 21 deg$^{2}$)  is somewhat lower than our expected average rate ($\sim\!62$) due to the slightly poorer IQ achieved in that part of the survey and the steepness of the \tno\ luminosity function. Subsequent blocks are being acquired with tighter attention to IQ limits to help ensure the anticipated discovery rate is achieved. 

\begin{figure}
\plotone{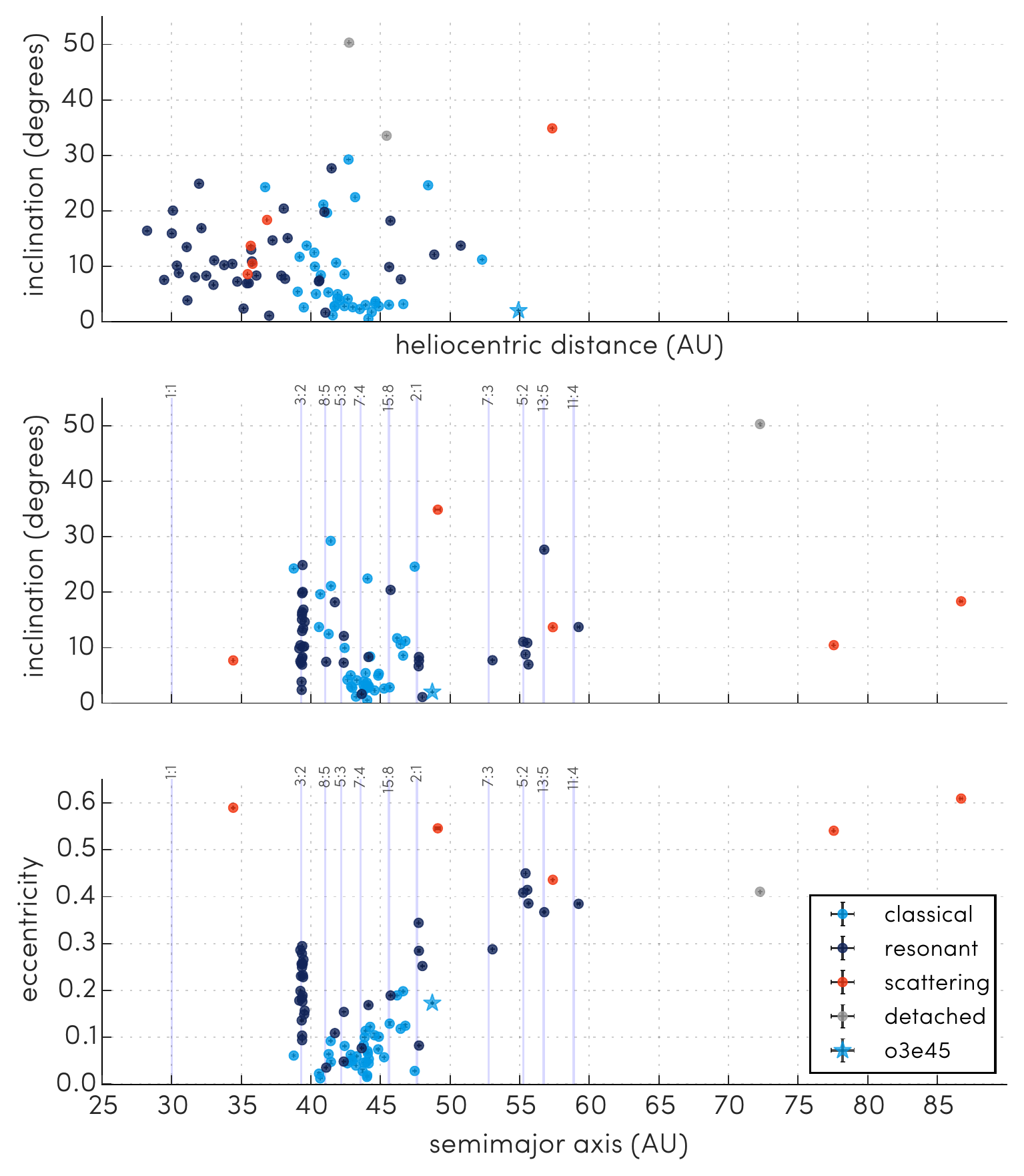}
\caption{Orbital parameters and discovery distances of the 85 characterized \ossos discoveries (Table~\ref{tab:discoveries}). 
\texttt{o3e01, o3o14, o3e39} are beyond the $a/e$ and $a/i$ axes ranges and are excluded from those two sub-plots for clarity.
These objects predominantly have orbital arcs of between 353 and 524 days (six have decade-long arcs).
The uncertainties are from the covariant matrix fit of \citet{Bernstein:2000p444};
note that they are sufficiently small that most error bars are smaller than the point size.
\texttt{o3e45} (2013 GQ$_{136}$) (star), a cold classical beyond the 2:1 resonance with Neptune, is discussed in \S~\ref{sec:tail}.
The pale blue vertical lines show the approximate semi-major axis locations of the resonance centers for resonances with \ossos detections.
\label{fig:orbdiscoveries}}
\end{figure}

\clearpage

\begin{deluxetable}{ccccccccccccc}
\tabletypesize{\scriptsize}
\rotate
\tablecolumns{13}
\tablecaption{Orbit and discovery properties of the characterized \ossos objects \label{tab:discoveries}}
\tablehead{\colhead{$m_{r}$} \vspace{-0.1cm} & \colhead{$\sigma$ $m_{r}$} & \colhead{Eff.} & \colhead{RA ($^{\circ}$)} & \colhead{Dec ($^{\circ}$)} & \colhead{a} & \colhead{e} & \colhead{i} & \colhead{Dist.} & \colhead{H$_{r}$} & \colhead{MPC} & \colhead{Object} & \colhead{Status} \\\colhead{discovery} & \colhead{all obs} & \colhead{} & \colhead{discov.} & \colhead{discov.} & \colhead{(AU)} & \colhead{} & \colhead{($^{\circ}$)} & \colhead{(AU)} & \colhead{} & \colhead{design.} & \colhead{} & \colhead{} }
\startdata 
\cutinhead{Centaurs}
23.39(6) & 0.17 & 0.78 & 239.535 & -12.008 & 22.144(2) & 0.37857(6) & 32.021(1) & 13.77(4) & 11.95 & 2013 JC$_{64}$ & o3o01 &       \\ 
\cutinhead{Inner classical belt} 
23.7(2) & 0.34 & 0.65 & 216.735 & -14.223 & 38.770(9) & 0.061(1) & 24.277(1) & 36.715(2) & 8.0 & 2013 GO$_{136}$ & o3e10 &       \\ 
\cutinhead{Main classical belt} 
22.97(9) & 0.21 & 0.78 & 210.435 & -10.419 & 44.10(1) & 0.066(2) & 2.762(1) & 41.714(2) & 6.7 & 2013 GN$_{137}$ & o3e22 &     I \\ 
22.99(5) & 0.14 & 0.78 & 214.785 & -11.817 & 45.259(4) & 0.05729(5) & 2.633(0) & 43.007(1) & 6.59 & 2013 EM$_{149}$ & o3e30PD &       \\ 
23.1(2) & 0.23 & 0.77 & 214.170 & -13.232 & 43.239(4) & 0.03952(6) & 1.171(0) & 41.569(1) & 6.82 & 2001 FK$_{185}$ & o3e20PD &       \\ 
23.1(1) & 0.14 & 0.77 & 213.630 & -11.944 & 44.153(4) & 0.04422(6) & 2.822(0) & 42.409(1) & 6.77 & 2004 EU$_{95}$ & o3e27PD &       \\ 
23.2(1) & 0.34 & 0.76 & 212.175 & -11.653 & 43.294(6) & 0.059(1) & 4.128(1) & 42.647(1) & 6.82 & 2013 GX$_{137}$ & o3e28 &       \\ 
23.26(9) & 0.31 & 0.75 & 216.525 & -13.051 & 47.459(5) & 0.028(1) & 24.608(2) & 48.424(1) & 6.34 & 2013 GG$_{138}$ & o3e44 &       \\ 
23.3(2) & 0.29 & 0.74 & 211.245 & -12.982 & 44.07(3) & 0.071(4) & 22.463(2) & 43.178(4) & 6.9 & 2013 GM$_{137}$ & o3e51 &       \\ 
23.37(9) & 0.18 & 0.73 & 214.845 & -13.373 & 43.963(3) & 0.04642(7) & 3.316(0) & 44.621(1) & 6.81 & 2004 HJ$_{79}$ & o3e37PD &       \\ 
23.37(8) & 0.31 & 0.73 & 215.040 & -11.853 & 46.447(3) & 0.11808(4) & 10.63(6) & 41.814(1) & 7.09 & 2001 FO$_{185}$ & o3e23PD &       \\ 
23.40(9) & 0.23 & 0.73 & 213.390 & -10.854 & 45.66(5) & 0.129(4) & 2.848(1) & 41.694(3) & 7.12 & 2013 GQ$_{137}$ & o3e21 &       \\ 
23.4(2) & 0.18 & 0.73 & 235.995 & -11.138 & 40.674(7) & 0.0122(8) & 19.641(2) & 41.166(4) & 7.23 & 2013 JN$_{65}$ & o3o28 &       \\ 
23.4(3) & 0.35 & 0.73 & 214.530 & -11.879 & 43.80(1) & 0.083(2) & 3.197(1) & 46.639(2) & 6.67 & 2013 GV$_{137}$ & o3e43 &     I \\ 
23.46(8) & 0.21 & 0.72 & 211.020 & -10.619 & 41.425(9) & 0.092(1) & 29.252(1) & 42.703(1) & 7.09 & 2013 GO$_{137}$ & o3e29 &       \\ 
23.5(1) & 0.37 & 0.72 & 214.080 & -11.197 & 43.864(5) & 0.09974(9) & 2.595(1) & 39.490(2) & 7.44 & 2013 GS$_{137}$ & o3e16 &       \\ 
23.5(1) & 0.34 & 0.71 & 212.325 & -11.247 & 43.717(8) & 0.028(4) & 1.748(1) & 44.371(2) & 6.94 & 2013 GP$_{137}$ & o3e35 &       \\ 
23.5(2) & 0.30 & 0.71 & 211.530 & -12.098 & 44.884(8) & 0.1010(7) & 5.309(1) & 41.250(1) & 7.29 & 2013 GY$_{137}$ & o3e53 &       \\ 
23.5(1) & 0.20 & 0.72 & 239.085 & -12.633 & 46.20(1) & 0.1893(6) & 11.707(1) & 39.188(1) & 7.53 & 2013 JR$_{65}$ & o3o21 &       \\ 
23.5(1) & 0.20 & 0.71 & 216.015 & -11.95 & 42.89(1) & 0.051(3) & 3.022(1) & 43.926(2) & 7.02 & 2013 GC$_{138}$ & o3e32 &       \\ 
23.6(1) & 0.26 & 0.70 & 214.560 & -14.305 & 44.58(4) & 0.104(4) & 2.294(2) & 43.515(2) & 7.1 & 2013 GT$_{137}$ & o3e31 &       \\ 
23.6(1) & 0.30 & 0.70 & 216.585 & -14.088 & 44.045(4) & 0.0187(1) & 0.551(0) & 44.130(1) & 7.05 & 2013 GF$_{138}$ & o3e34PD &       \\ 
23.59(9) & 0.37 & 0.69 & 214.710 & -13.957 & 44.837(9) & 0.074(1) & 4.973(1) & 41.922(2) & 7.3 & 2013 GU$_{137}$ & o3e25 &       \\ 
23.6(1) & 0.21 & 0.69 & 215.805 & -12.522 & 42.975(5) & 0.0499(6) & 2.787(1) & 44.882(1) & 7.01 & 2013 GB$_{138}$ & o3e38 &       \\ 
23.6(1) & 0.24 & 0.69 & 211.920 & -11.691 & 42.862(4) & 0.0625(3) & 5.017(1) & 40.370(1) & 7.5 & 2013 GW$_{137}$ & o3e54 &       \\ 
23.8(2) & 0.27 & 0.61 & 216.465 & -14.855 & 44.17(2) & 0.053(3) & 3.069(2) & 45.611(2) & 7.16 & 2013 GE$_{138}$ & o3e40 &       \\ 
23.8(1) & 0.36 & 0.61 & 215.460 & -12.938 & 44.027(9) & 0.0153(9) & 3.677(2) & 44.638(2) & 7.3 & 2013 HT$_{156}$ & o3e36 &       \\ 
23.8(2) & 0.22 & 0.60 & 215.700 & -12.322 & 43.800(4) & 0.0458(6) & 3.900(1) & 42.097(1) & 7.52 & 2013 GA$_{138}$ & o3e26 &       \\ 
23.9(1) & 0.36 & 0.58 & 216.045 & -12.328 & 43.931(5) & 0.1136(3) & 5.421(1) & 39.054(1) & 7.87 & 2013 GD$_{138}$ & o3e15 &       \\ 
23.9(4) & 0.26 & 0.58 & 215.460 & -12.182 & 41.44(2) & 0.047(4) & 21.117(2) & 40.894(2) & 7.69 & 2013 GZ$_{137}$ & o3e18 &       \\ 
23.89(9) & 0.21 & 0.68 & 236.085 & -12.921 & 46.79(2) & 0.124(3) & 11.206(2) & 52.299(4) & 6.67 & 2013 JM$_{65}$ & o3o35 &       \\ 
24.0(1) & 0.41 & 0.47 & 213.885 & -12.325 & 42.63(1) & 0.044(4) & 4.226(1) & 41.903(2) & 7.72 & 2013 GR$_{137}$ & o3e24 &       \\ 
24.0(1) & 0.14 & 0.66 & 236.370 & -10.629 & 41.278(6) & 0.0639(9) & 12.468(1) & 40.245(1) & 7.93 & 2013 JP$_{65}$ & o3o23 &       \\ 
24.1(1) & 0.20 & 0.64 & 238.725 & -11.913 & 44.26(1) & 0.122(1) & 8.413(0) & 40.716(2) & 8.0 & 2013 JQ$_{65}$ & o3o26 &       \\ 
24.2(1) & 0.19 & 0.64 & 241.410 & -11.576 & 40.58(1) & 0.022(2) & 13.729(1) & 39.698(3) & 8.12 & 2013 JS$_{65}$ & o3o22 &       \\ 
24.2(1) & 0.16 & 0.63 & 242.130 & -12.475 & 46.63(3) & 0.198(1) & 8.573(0) & 42.413(2) & 7.84 & 2013 JT$_{65}$ & o3o30 &       \\ 
24.4(2) & 0.19 & 0.57 & 236.085 & -10.613 & 42.42(1) & 0.0814(9) & 9.958(1) & 40.290(2) & 8.26 & 2013 JO$_{65}$ & o3o24 &       \\ 
\cutinhead{Outer classical belt} 
23.6(1) & 0.24 & 0.69 & 215.070 & -13.474 & 48.72(2) & 0.173(2) & 2.031(2) & 54.915(2) & 6.13 & 2013 GQ$_{136}$ & o3e45 &       \\ 
\cutinhead{Detached classical belt} 
23.07(7) & 0.19 & 0.77 & 211.890 & -11.161 & 149.8(5) & 0.726(1) & 33.539(1) & 45.442(1) & 6.42 & 2013 GP$_{136}$ & o3e39 &     I \\ 
24.4(2) & 0.23 & 0.55 & 240.510 & -11.985 & 72.26(2) & 0.4105(2) & 50.318(1) & 42.745(2) & 8.01 & 2013 JD$_{64}$ & o3o31 &       \\ 
\cutinhead{Objects in resonance with Neptune} 
22.69(7) & 0.22 & 0.81 & 216.270 & -14.536 & 47.74(2) & 0.3440(4) & 6.660(1) & 33.001(1) & 7.42 & 2013 GW$_{136}$ & o3e05 & 2:1   \\ 
23.4(1) & 0.36 & 0.73 & 211.845 & -12.285 & 48.01(1) & 0.2519(5) & 1.100(1) & 37.002(1) & 7.67 & 2013 GX$_{136}$ & o3e55 & 2:1   \\ 
23.6(1) & 0.15 & 0.72 & 236.655 & -13.161 & 47.76(6) & 0.284(2) & 8.335(1) & 36.086(2) & 7.94 & 2013 JE$_{64}$ & o3o18 & 2:1   \\ 
24.0(1) & 0.24 & 0.67 & 237.390 & -12.517 & 47.77(1) & 0.082(1) & 7.65(0) & 46.465(2) & 7.27 & 2013 JJ$_{64}$ & o3o33 & 2:1   \\ 
\hline
21.15(2) & 0.09 & 0.85 & 236.775 & -11.987 & 39.36(5) & 0.184(3) & 15.081(1) & 38.330(2) & 5.27 & 2007 JF$_{43}$ & o3o20PD & 3:2   \\ 
23.23(6) & 0.13 & 0.75 & 237.645 & -13.115 & 39.403(4) & 0.18887(8) & 24.898(1) & 31.965(1) & 8.13 & 2013 JB$_{65}$ & o3o09 & 3:2   \\ 
23.3(1) & 0.21 & 0.74 & 213.840 & -13.5 & 39.44(1) & 0.2282(7) & 13.468(1) & 31.080(1) & 8.32 & 2013 GH$_{137}$ & o3e02 & 3:2   \\ 
23.4(2) & 0.27 & 0.73 & 214.695 & -11.658 & 39.47(3) & 0.265(1) & 16.873(1) & 32.135(1) & 8.25 & 2013 GJ$_{137}$ & o3e04 & 3:2   \\ 
23.40(8) & 0.16 & 0.73 & 240.945 & -11.399 & 39.37(2) & 0.2555(9) & 19.815(1) & 40.970(1) & 7.22 & 2013 JJ$_{65}$ & o3o27 & 3:2   \\ 
23.48(7) & 0.25 & 0.72 & 237.360 & -11.28 & 39.371(4) & 0.0937(1) & 13.015(1) & 35.715(1) & 7.9 & 2013 JD$_{65}$ & o3o15 & 3:2   \\ 
23.62(8) & 0.19 & 0.71 & 238.020 & -12.35 & 39.363(5) & 0.2493(2) & 15.934(1) & 30.010(1) & 8.79 & 2013 JG$_{65}$ & o3o04 & 3:2   \\ 
23.67(7) & 0.22 & 0.71 & 237.225 & -11.123 & 39.375(5) & 0.2944(2) & 16.409(1) & 28.231(1) & 9.11 & 2013 JC$_{65}$ & o3o02 & 3:2   \\ 
23.69(9) & 0.21 & 0.70 & 236.145 & -10.369 & 39.419(6) & 0.2326(1) & 10.12(8) & 30.375(1) & 8.82 & 2013 JZ$_{64}$ & o3o06 & 3:2   \\ 
23.7(1) & 0.30 & 0.65 & 211.890 & -13.064 & 39.33(3) & 0.257(1) & 3.866(1) & 31.131(1) & 8.7 & 2013 GE$_{137}$ & o3e03 & 3:2   \\ 
23.9(1) & 0.45 & 0.56 & 212.805 & -12.83 & 39.56(1) & 0.1567(8) & 14.680(1) & 37.246(1) & 8.11 & 2013 GF$_{137}$ & o3e12 & 3:2   \\ 
23.9(1) & 0.29 & 0.52 & 216.690 & -13.261 & 39.17(1) & 0.178(1) & 9.879(2) & 45.625(2) & 7.28 & 2013 GK$_{137}$ & o3e41 & 3:2   \\ 
24.0(1) & 0.23 & 0.67 & 239.265 & -12.607 & 39.24(3) & 0.286(1) & 7.553(0) & 29.458(1) & 9.22 & 2013 JH$_{65}$ & o3o03 & 3:2   \\ 
24.0(1) & 0.66 & 0.45 & 210.960 & -11.526 & 39.371(5) & 0.1035(5) & 6.943(1) & 35.413(1) & 8.45 & 2013 GD$_{137}$ & o3e08 & 3:2   \\ 
24.0(3) & 0.32 & 0.44 & 217.395 & -13.633 & 39.25(3) & 0.199(2) & 10.440(1) & 34.356(1) & 8.59 & 2013 GL$_{137}$ & o3e06 & 3:2   \\ 
24.1(2) & 0.33 & 0.41 & 212.640 & -10.849 & 39.34(2) & 0.136(2) & 2.391(1) & 35.160(1) & 8.52 & 2013 GG$_{137}$ & o3e07 & 3:2   \\ 
24.1(1) & 0.21 & 0.65 & 238.605 & -13.216 & 39.389(4) & 0.1762(1) & 8.316(0) & 32.482(1) & 8.94 & 2013 JF$_{65}$ & o3o10 & 3:2   \\ 
24.1(2) & 0.18 & 0.65 & 236.910 & -10.624 & 39.520(5) & 0.1488(2) & 10.223(0) & 33.771(1) & 8.78 & 2013 JA$_{65}$ & o3o12 & 3:2   \\ 
24.2(1) & 0.20 & 0.63 & 238.305 & -13.255 & 39.358(9) & 0.2784(4) & 8.048(0) & 31.676(1) & 9.15 & 2013 JE$_{65}$ & o3o08 & 3:2   \\ 
24.3(1) & 0.28 & 0.61 & 243.045 & -13.607 & 39.29(1) & 0.2306(7) & 7.251(0) & 34.714(1) & 8.79 & 2013 JL$_{65}$ & o3o13 & 3:2   \\ 
24.3(1) & 0.24 & 0.60 & 241.455 & -12.778 & 39.416(6) & 0.2566(2) & 20.045(1) & 30.089(1) & 9.44 & 2013 JK$_{65}$ & o3o05 & 3:2   \\ 
\hline
22.94(4) & 0.17 & 0.77 & 238.425 & -12.45 & 55.250(9) & 0.4083(1) & 11.077(0) & 33.054(1) & 7.69 & 2013 JK$_{64}$ & o3o11 & 5:2   \\ 
22.94(5) & 0.19 & 0.79 & 216.855 & -15.028 & 55.55(3) & 0.4143(5) & 10.877(1) & 35.765(1) & 7.32 & 2013 GY$_{136}$ & o3e09 & 5:2   \\ 
23.9(3) & 0.20 & 0.68 & 236.805 & -12.989 & 55.42(1) & 0.44971(9) & 8.785(0) & 30.514(1) & 8.97 & 2013 JF$_{64}$ & o3o07 & 5:2   \\ 
23.9(2) & 0.27 & 0.54 & 210.480 & -10.686 & 55.63(3) & 0.3855(6) & 6.978(1) & 35.539(2) & 8.34 & 2013 GS$_{136}$ & o3e48 & 5:2   \\ 
\hline
24.1(2) & 0.22 & 0.40 & 211.260 & -10.733 & 42.370(4) & 0.1540(2) & 12.112(2) & 48.863(2) & 7.11 & 2013 GT$_{136}$ & o3e52 & 5:3 IH \\ 
24.1(1) & 0.29 & 0.64 & 242.025 & -13.547 & 42.358(5) & 0.0481(5) & 7.287(0) & 40.573(1) & 7.98 & 2013 JM$_{64}$ & o3o25 & 5:3 I \\ 
\hline
23.8(2) & 0.24 & 0.69 & 242.010 & -12.902 & 53.05(1) & 0.2876(3) & 7.74(0) & 38.148(1) & 7.96 & 2013 JN$_{64}$ & o3o19 & 7:3   \\ 
23.4(1) & 0.35 & 0.73 & 211.185 & -12.217 & 43.649(7) & 0.0767(7) & 1.645(1) & 41.043(1) & 7.2 & 2013 GR$_{136}$ & o3e19 & 7:4   \\ 
24.0(1) & 0.20 & 0.47 & 214.920 & -13.83 & 41.100(6) & 0.035(1) & 7.452(1) & 40.609(1) & 7.85 & 2013 GV$_{136}$ & o3e17 & 8:5 IH \\ 
22.70(4) & 0.11 & 0.79 & 237.195 & -13.113 & 59.23(8) & 0.385(2) & 13.731(1) & 50.766(2) & 5.6 & 2013 JH$_{64}$ & o3o34 & 11:4 I \\ 
23.3(1) & 0.23 & 0.75 & 238.815 & -12.604 & 56.77(5) & 0.367(1) & 27.672(1) & 41.487(1) & 7.03 & 2013 JL$_{64}$ & o3o29 & 13:5 IH \\ 
23.54(9) & 0.22 & 0.70 & 215.355 & -12.892 & 45.73(1) & 0.1889(5) & 20.412(1) & 38.052(1) & 7.72 & 2013 HR$_{156}$ & o3e49 & 15:8 I \\ 
23.7(1) & 0.31 & 0.65 & 212.220 & -10.496 & 44.14(2) & 0.169(1) & 8.318(1) & 37.876(1) & 7.86 & 2013 GU$_{136}$ & o3e13 & 16:9 IH \\ 
24.1(1) & 0.20 & 0.64 & 236.670 & -10.521 & 41.725(7) & 0.1088(7) & 18.208(1) & 45.715(2) & 7.48 & 2013 JG$_{64}$ & o3o32 & 18:11 IH \\ 
\cutinhead{Scattering disk} 
21.50(9) & 0.18 & 0.88 & 213.150 & -13.587 & 34.42(4) & 0.5897(6) & 7.711(1) & 23.291(1) & 7.73 & 2002 GG$_{166}$ & o3e01 &       \\ 
23.54(8) & 0.17 & 0.72 & 237.030 & -12.827 & 143.31(9) & 0.7548(2) & 8.58(0) & 35.456(1) & 8.0 & 2013 JO$_{64}$ & o3o14 &       \\ 
23.6(1) & 0.65 & 0.69 & 210.615 & -12.965 & 86.72(9) & 0.6092(5) & 18.363(1) & 36.851(1) & 7.86 & 2013 GZ$_{136}$ & o3e11 &       \\ 
23.73(9) & 0.17 & 0.70 & 241.740 & -14.215 & 49.1(2) & 0.546(3) & 34.876(4) & 57.339(6) & 6.09 & 2013 JQ$_{64}$ & o3o36 &     I \\ 
23.9(1) & 0.15 & 0.68 & 239.805 & -12.537 & 57.38(4) & 0.4359(6) & 13.701(1) & 35.680(1) & 8.34 & 2013 JP$_{64}$ & o3o16 &       \\ 
24.3(1) & 0.32 & 0.59 & 241.440 & -12.657 & 77.57(2) & 0.5406(2) & 10.459(1) & 35.811(1) & 8.71 & 2013 JR$_{64}$ & o3o17 &       \\ 
\enddata 
\tablecomments{
Numbers in parentheses are the uncertainty in the last given digit.
$m_{r}$ discovery is an average magnitude during the discovery triplet only, eliding any measurements with photometric flags.
$\sigma m_{r}$ is the standard deviation of all measured magnitudes without photometric flags.
Eff. is the value of the detection efficiency function for the motion rate and magnitude of the object at its discovery.
$a, e, i$ are J2000 ecliptic barycentric coordinates of semi-major axis, eccentricity, inclination,
with uncertainties from the covariant matrix fit of \citet{Bernstein:2000p444};
full barycentric elements are available at \url{http://www.ossos-survey.org/}.
The full heliocentric orbital elements are available in electronic form from the Minor Planet Center.
We assign survey designations here based on their \ossos discovery,
with a format $o$ for \textsc{ossos}, the last digit of the year in which the object was discovered by \ossos
 (3-6), the block ID letter ($e, o$) and the sequential number 01-xx to give unique identifiers.
``PD" indicates previous discovery.
$p:q$: object is in the $p:q$ resonance;
I: the orbit classification is currently insecure;
H: the human operator intervened to declare the orbit security status.
}
\end{deluxetable} 

\begin{deluxetable}{cccccccccccc}
\tabletypesize{\scriptsize}
\rotate
\tablecolumns{12}
\tablecaption{Orbit and discovery properties of the uncharacterized \ossos objects \label{tab:unchar_discoveries}}
\tablehead{\colhead{$m_{r}$} \vspace{-0.1cm} & \colhead{$\sigma$ $m_{r}$} & \colhead{Eff.} & \colhead{RA ($^{\circ}$)} & \colhead{Dec ($^{\circ}$)} & \colhead{a} & \colhead{e} & \colhead{i} & \colhead{Dist.} & \colhead{H$_{r}$} & \colhead{MPC} & \colhead{Object}  \\\colhead{discovery} & \colhead{all obs} & \colhead{} & \colhead{discov.} & \colhead{discov.} & \colhead{(AU)} & \colhead{} & \colhead{($^{\circ}$)} & \colhead{(AU)} & \colhead{} & \colhead{design.} & \colhead{} }
\startdata 
\cutinhead{Arc outside discovery lunation (via serendipitous tracking observations)}
24.1(1) & 0.21 & 0.35 & 215.790 & -13.635 & 45.65(4) & 0.176(2) & 8.261(1) & 38.536(3) & 8.23 & 2013 HS$_{156}$ & uo3e14  \\ 
24.3(2) & 0.44 & 0.15 & 214.380 & -13.558 & 43.22(1) & 0.024(3) & 2.308(2) & 43.812(2) & 7.81 & 2013 GC$_{137}$ & uo3e33   \\ 
24.5(2) & 0.26 & 0.45 & 237.615 & -11.245 & 40.179(9) & 0.0702(6) & 26.609(3) & 37.372(3) & 8.72 & 2013 JU$_{64}$ & uo3o37  \\ 
24.5(3) & 0.25 & 0.43 & 236.640 & -12.721 & 42.527(8) & 0.049(8) & 9.633(2) & 40.645(5) & 8.39 & 2013 JS$_{64}$ & uo3o38  \\ 
24.6(2) & 0.31 & 0.36 & 238.125 & -11.727 & 73.0(7) & 0.575(6) & 31.388(2) & 53.286(3) & 7.26 & 2013 JV$_{64}$ & uo3o50  \\ 
\cutinhead{Arc only within discovery lunation}
24.2(2) & 0.29 & 0.27 &  212.565 & -11.181 & 43(22)  & 0.0(6) & 12.2(4.7) & 42.2(2.7) & 7.85 & & uo3e50nt  \\ 
24.4(1) & 0.05 & 0.55 & 240.285 & -13.07 & 44(23)  & 0.0(6) & 27(26)  & 43.0(5.1) & 8.0 & & uo3o46nt  \\ 
24.4(2) & 0.02 & 0.53 & 237.975 & -11.892 & 38(20)  & 0.0(6) & 9.6(9) & 37.4(2.4) & 8.64 & & uo3o43nt  \\ 
24.5(2) & 0.23 & 0.46 & 241.665 & -12.826 & 35(19)  & 0.0(6) & 11.2(1.9) & 34.3(2.5) & 9.07 & & uo3o39nt  \\ 
24.5(2) & 0.10 & 0.45 & 237.300 & -12.453 & 42(22)  & 0.0(6) & 23(24)  & 41.3(4.9) & 8.29 & & uo3o47nt  \\ 
24.5(2) & 0.25 & 0.04 & 212.340 & -10.629 & 50(25)  & 0.3(5) & 2.6(1) & 44.7(1.3) & 7.93 & & uo3e42  \\ 
24.5(2) & 0.17 & 0.44 & 241.350 & -13.576 & 36(19)  & 0.0(6) & 11(13)  & 35.1(3.7) & 9.0 & & uo3o40nt  \\ 
24.5(2) & 0.19 & 0.44 & 241.650 & -13.872 & 38(20)  & 0.0(6) & 9(12)  & 37.1(3.9) & 8.75 & & uo3o42nt  \\ 
24.6(2) & 0.26 & 0.01 & 213.645 & -11.146 & 16(12)  & 0.1(9) & 32(32)  & 14.5(4.6) & 12.88 & & uo3e46nt  \\ 
24.6(2) & 0.14 & 0.34 & 241.800 & -13.145 & 37(20)  & 0.0(6) & 9.0(9) & 36.0(2.5) & 8.96 & & uo3o41nt  \\ 
24.7(2) & -- & 0.23 & 238.695 & -11.017 & 40(21)  & 0.0(6) & 8.9(1.9) & 38.7(4.1) & 8.74 & & uo3o44nt  \\ 
24.7(2) & 0.23 & 0.19 & 242.460 & -13.232 & 44(23)  & 0.0(6) & 7.8(5) & 42.7(2.6) & 8.33 & & uo3o45nt  \\ 
24.7(2) & -- & 0.19 & 237.000 & -12.451 & 47(25)  & 0.0(6) & 40(49)  & 46.5(8.4) & 7.98 & & uo3o48nt  \\ 
24.8(3) & -- & 0.11 & 236.865 & -10.903 & 49(25)  & 0.0(6) & 8.9(6.2) & 48.4(5.5) & 7.89 & & uo3o49nt  \\ 
\enddata 
\tablecomments{
Serendipitous tracking observations are those where the object happened to be within and was visible in images taken to extend the orbital arc of characterized \ossos objects.
Numbers in parentheses are the uncertainty in the last given digit.
$m_{r}$ discovery is an average magnitude during the discovery triplet only, eliding any measurements with photometric flags. 
$\sigma$ $m_{r}$ is the standard deviation of all measured magnitudes without photometric flags.
Eff. is the value of the detection efficiency function for the motion rate of the object at its discovery.
$a, e, i$ are J2000 ecliptic barycentric coordinates of semi-major axis, eccentricity, inclination,
with uncertainties from the covariant matrix fit of \citet{Bernstein:2000p444};
full barycentric elements are available at \url{http://www.ossos-survey.org/}.
The full heliocentric orbital elements are available in electronic form from the Minor Planet Center.
We assign survey designations here based on their \ossos discovery,
with a format $o$ for \textsc{ossos}, the last digit of the year in which the object was discovered by \ossos
 (3-6), the block ID letter ($e, o$) and the sequential number 01-xx to give unique identifiers.
Orbital classifications are not applied to these objects.
``u'' indicates uncharacterized; ``nt'' at the end of the object ID designates not observed on more than two nights in the discovery lunation.
}
\end{deluxetable}

\section{Substructure of the classical belt}
\label{sec:classical}

\citet{Petit:2011p3938} noted the need for substructure in the main classical belt. This is the non-resonant population with semi-major axes between the 3:2 and 2:1 mean-motion resonances with Neptune, though considering only $40 \leq a \leq 47$ \textsc{au} to avoid the complex resonance boundaries around 39.4 and 47.8 \au respectively.
That work showed that the main classical belt could be modelled with three probability distributions within $a/e/i$ phase space. (Hereafter: the \cfeps {\it L7 model}; see Fig. 4 and Appendix A in \citet{Petit:2011p3938}). With the first \ossos sample, we confirm this three-component view.

We illustrate our three-component model in Fig.~\ref{fig:qi}.
We describe the dynamically excited hot classical belt with a single smooth {\it hot component}: with width $\sigma_h = 16^{\circ}$ in inclination (most visible at $i > 7^{\circ}$), continuously covering all stable semi-major axes $a$ beyond Neptune. We impose a void on the model in the region $i < 12^{\circ}$, $a < 42.4$ \au to account for the destabilizing action of the $\nu_{8}$ secular resonance (Fig.~\ref{fig:qi}, lightest grey points).
The cold classical belt is described by a low-inclination band that begins beyond $a=42.5$~\textsc{au}, with inclination width of roughly $2^\circ$, most visible at $i < 7^{\circ}$ (Fig.~\ref{fig:qi}, lower left, darker grey points). This cold belt is created by a superposition of two components, which are termed the \textit{kernel} and the \textit{stirred} components: these are discussed in detail in \S~\ref{sec:kernel}.

Considering the perihelion distribution in the main classical belt,
we also confirm the difference in the perihelion distribution of the hot
and cold main-belt populations seen by \citet{Petit:2011p3938} (Fig.~\ref{fig:qi}, upper right).
The hot population seen by \ossos is concentrated in the perihelion range $q=$ 35--41~\textsc{au}, with soft exponential decay about an \au to either side, while the cold belt population has perihelia 38--47~\textsc{au} (Fig.~\ref{fig:qi}).

Interior to the main belt, the inner classical belt objects inhabit a more limited stable phase space, due to the $\nu_{8}$ secular resonance. 
The inner belt here comprise the non-resonant, non-scattering population $a_{Neptune} < a < $ 3:2 mean-motion resonance. 
Inner belt objects detected in previous surveys in the $a=34$--39~\au range are consistent with being detections of the low-$a$
tail of the main belt hot population \citep{Petit:2011p3938}, based on applying the \cfeps survey simulator.
Photometric studies support this conclusion through colours more consistent with those of the hot main-belt population rather than the distinctly red cold classicals \citep{Peixinho:2015bw, Romanishin:2010we}.
The lone \ossos detection in the inner classical belt, \texttt{o3e10}, has $i = 24^{\circ}$ (Fig.~\ref{fig:qi}, lower left).
Using the \cfeps survey simulator, the sample therefore remains consistent with the inner classical belt being a lower-$a$ tail of the hot main belt.


\begin{figure}
\plotone{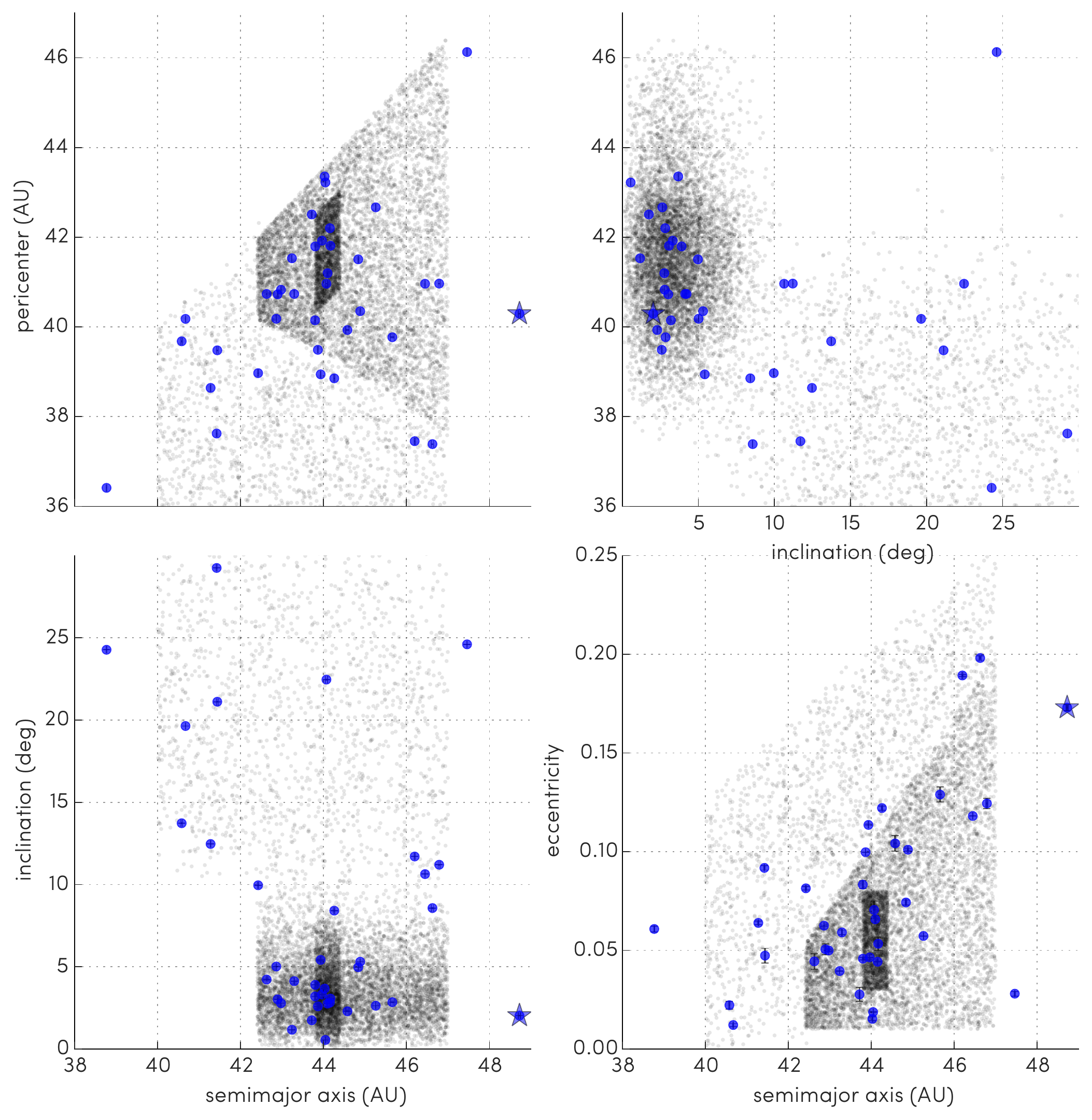}
\caption{An illustration of the three-component L7 model \citep{Petit:2011p3938} model for the structure of the classical belt, overlaid by the classical belt as observed by \textsc{ossos}. 
Blue points are characterized \ossos discoveries that are classified in the classical belt (Table~\ref{tab:discoveries}). 
Outer classical belt object \texttt{o3e45} (star) is discussed in \S~\ref{sec:tail}. 
The L7 model has three probability density functions that together describe the intrinsic classical main-belt population for the 40-47 \au region between the 3:2 and 2:1 resonances (\S~\ref{sec:classical}): the dynamically excited hot component (lightest grey), the dynamically stirred cold main belt (mid grey), and the dynamically quiescent cold classical kernel (darkest grey). 
\label{fig:qi}}
\end{figure}

\subsection{A kernel exists in the cold classical belt}
\label{sec:kernel}

One of the important findings of \citet{Petit:2011p3938} was the substructure
present in the $a/e$ distribution of the cold component of the main classical belt. 
With the \ossos first-quarter sample, we can test for the existence of this structure in an independent data set.
\citet{Petit:2011p3938} represented this substructure in the L7 model by two superposed components. 
A small {\it kernel component} compact in $a$ and in $e$ was centered near 44~\textsc{au}.
Overlaying this was a second population, the {\it stirred component},
which is smooth in semi-major axis distribution,
low in inclination, and occupies $q = 38$--44 \au non-uniformly, with $a = 42.4$--47~\textsc{au}.
Its inner edge begins at the $\nu _{8}$ secular resonance and the outer
bound is the 2:1 mean-motion resonance.
The stirred component could have been slightly dynamically agitated by weak interactions.
The split to two components was informed by the clumped, $a$-dependent nature of the $e$ distribution.
Fig.~\ref{fig:qi}, particularly the $a/i$ plot at the lower left, 
shows that indeed the over-density near $a=44$~\au is also present in the \ossos discoveries. 
 
However, to investigate if \citet{Petit:2011p3938} over-interpreted the previous detections,
we tested the detected \ossos $a$ distribution in turn against a smooth distribution and against 
the L7 model of the classical belt substructure (Fig.~\ref{fig:nokernel}), 
using the same Anderson-Darling tests\footnote{\url{http://www.itl.nist.gov/div898/handbook/eda/section3/eda35e.htm}} 
for the $a$ distribution as were done by \citet{Petit:2011p3938}. 
The data demand a substructure in the cold component:
a model using only a smooth $a$ distribution for the cold component, with no kernel,
was rejected at more than 95\% confidence by the \ossos detections.
We therefore confirm that there is a real ``kernel" concentration in the Kuiper belt in a narrow semi-major
axis range around 44~\textsc{au}.
While it is plausible that other two-component models might be
used to represent the classical belt, the L7 model at present still provides
a valid representation of the orbital distribution for the main-belt's cold
component: it could not be rejected by the \ossos sample (Fig.~\ref{fig:nokernel}). 
However, the greater sample density of the kernel will provide further constraints on scenarios where 
the kernel formed as a fossil population from the former location of the 2:1 resonance, 
left as an effect of a discontinuous change of Neptune's semi-major axis during its migration \citep{Nesvorny:2015ik}.

\begin{figure}
\plotone{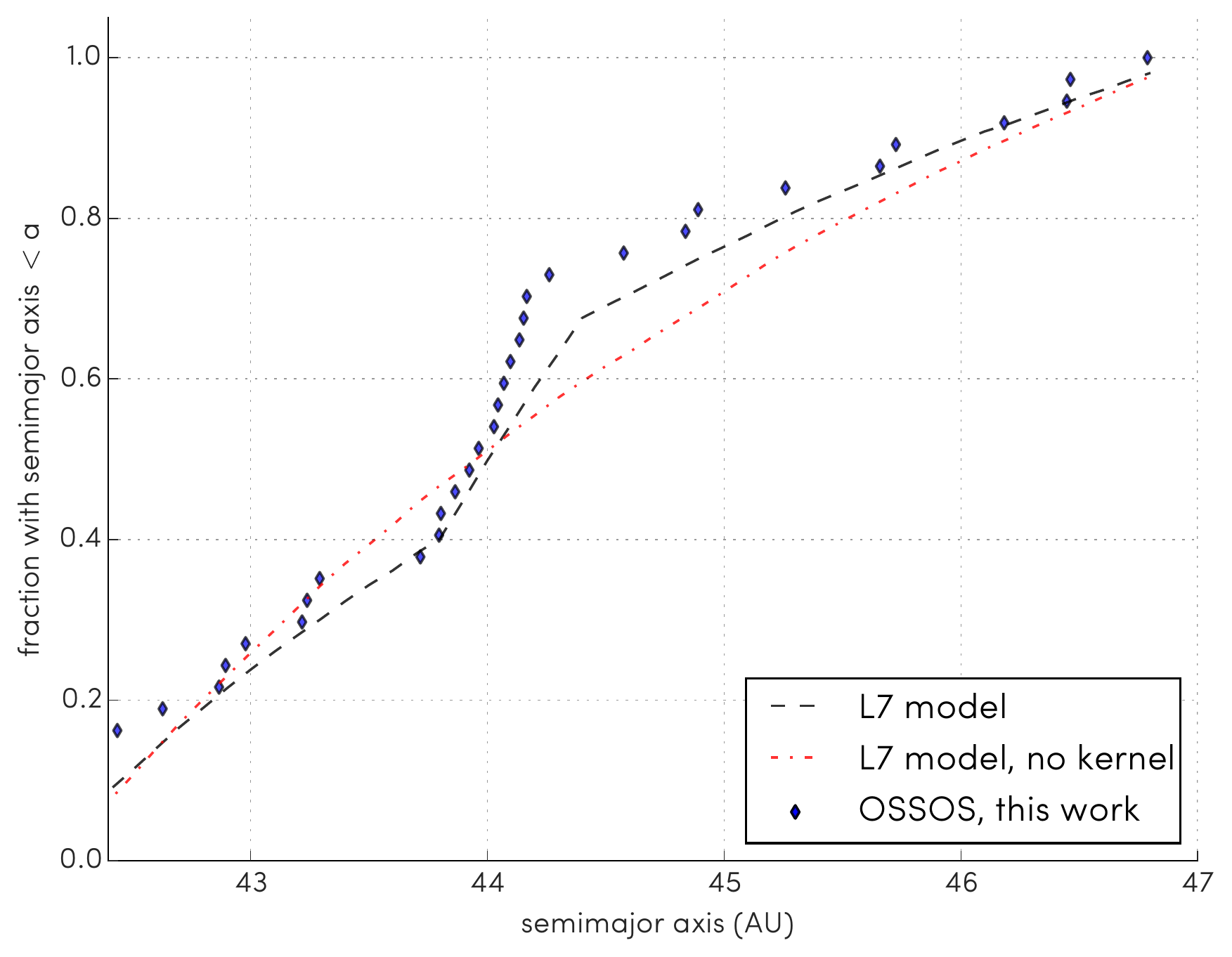}
\caption{Cumulative semi-major axis distribution of the first-quarter \ossos main-belt detections (diamonds).
The dashed curve shows the cumulative distribution of the
expected detections if the \cfeps L7 model of \citet{Petit:2011p3938} was the Solar System observed by \textsc{ossos},
as determined via the \ossos survey simulator. This model reasonably predicts the
high density of \ossos detections near 44~\textsc{au}, via a `kernel' subcomponent in the model.
Removing the kernel, and simulating the main-belt detections with a cold component 
that is instead purely smooth, produces a predicted semi-major axis distribution for the detections 
(dotted line) that is rejected at more than 95\% tolerance.
\label{fig:nokernel}}
\end{figure}

\subsection{A stirred tail of cold classicals beyond the 2:1 resonance?}
\label{sec:tail}

The first-quarter \ossos sample includes the newly discovered object \texttt{o3e45} (2013 GQ$_{136}$), which has
$a=48.72$~\textsc{au}, $e=0.173$, and $i=2.031^\circ$.
With $q=40.3$~\au (Fig.~\ref{fig:qi}), this object lies along a natural extension of the
stirred component. 
Crucially, its orbit is beyond the current barrier of the 2:1 resonance (Fig~\ref{fig:orbdiscoveries}).
If this object is part of the smoothly $a$-distributed `stirred' component that
we modelled in the cold main belt, there would be strong cosmogonic implications.
\texttt{o3e45} joins only a few other published objects with low-$i$ just beyond the 2:1 resonance (Table~\ref{tab:stirred}): 
particularly (48639) 1995 TL$_{18}$ \citep{2002Icar..157..269G}, 2003 UY$_{291}$ \citep{Gladman2008}, and 2011 US$_{412}$ \citep{Alexandersen:2014uv}.
The key structural features in this region are the 40--42~\au range where the kernel perihelia centre, 
the $a \simeq44.5$ \au outer edge of the kernel, and the 2:1 resonance, centred at 47.7~\textsc{au}, with a width $\pm0.4$~\textsc{au}.
In this context, these three objects imply a scenario where present $a>44.5$~\au members of the stirred
component are objects shifted from a primordial $a<44.5$~\textsc{au}.
In a past where Neptune's eccentricity was larger than at present, these objects were
stirred by gentle close encounters, which minimally modified their eccentricities and nudged them out into a higher-$a$ population tail.
Could there now be a continuous distribution of primordial cold objects, scattered
from initial $a = $ 40--42~\au orbits, that presently orbit with perihelia in at their original position? 
This could require cosmogonic models to scatter cold objects into a structure that reaches even beyond the 2:1 resonance, 
while creating or preserving a concentration of the same cold objects at $a \sim\! 44$~\textsc{au}.
Alternatively, these low-$i$ $a > 48$ \au objects could be in-situ remnants of an original disk extending to at least 50 \au \citep{Lykawka:2008ia}.
Their low number implies a relatively small population \citep{Nesvorny:2015ik}.

\begin{deluxetable}{crrrrrrc}
\tablecolumns{8}
\tablecaption{Low-inclination ($i < 5^{\circ}$) objects with $q > 40$ AU and $a$ beyond the 2:1 resonance, listed by the Minor Planet Center as of Feb 2016 \label{tab:stirred}}
\tablehead{\colhead{Object} & \colhead{i} & \colhead{e} & \colhead{q} & \colhead{a} & \colhead{Q} & \colhead{H} & \colhead{Comment} \\
\colhead{} & \colhead{ ($^{\circ}$)} & \colhead{} & \colhead{(AU)} & \colhead{(AU)} & \colhead{(AU)} & \colhead{(mag)} & \colhead{}
 }
\startdata
\cutinhead{``Stirred" objects with high orbital precision}
(48639) (1995 TL$_{8}$)	            &0.2 &	0.23 & 40.12 &	52.40 &	64.68 &	5.4   & \citet{2002Icar..157..269G} noted $q > 40$ \\
2003 UY$_{291}$ &	3.5 &	0.16 	& 41.35 &	49.28 &	57.21 &	7.4  & Identified in \citet{Gladman2008} \\
2011 US$_{412}$ & 2.6 & 0.16 & 40.03 & 47.76 & 55.48 &  7.7 & \citet{Alexandersen:2014uv}; not resonant \\
2013 GQ$_{136}$ &2.0 &	0.17 	& 40.63 &	48.87 &57.10 &	6.1 & This work: \texttt{o3e45} \\
\cutinhead{Large orbital uncertainties (unclassifiable) or poor orbit sampling}
2001 FL$_{193}$  &1.0 &	0.20 & 40.23 &	50.17 &	60.10 &	8.7  &  \\
2002 CP$_{154}$ &	1.5 &	0.20 	& 42.07 &	52.64 &	63.21 &	6.5  &   \\
2006 AO$_{101}$ &	1.1 &	0.21 	& 41.92 &	52.92 &	63.93 &	7.1 &   \\
\enddata
\tablecomments{
Heliocentric orbital elements from the MPC.
2001 FL$_{193}$ on its discovery in 2001 (MPEC 2001-U19: \url{http://www.minorplanetcenter.net/mpec/K01/K01U19.html}) was assigned to an $a = 44$ \textsc{au}, $e = 0.09$ orbit and subsequently lost. 
Its June 2015 recovery (MPEC 2015-M50: \url{http://www.minorplanetcenter.net/mpec/K15/K15M50.html}) revised its orbit to $a = 50.2$ \textsc{au}, $e = 0.20$. This echoes the perils of ephemeris bias \citep{Jones:2006jl}.}
\end{deluxetable}

The L7 model, which we confirmed in \S~\ref{sec:kernel}, did not have an `outer' cold main classical Kuiper belt beyond 47~\textsc{au}, as 
the hot classical component of the L7 model sufficiently explained the \cfeps detections.  
We therefore test if the stirred component of the L7 model of the cold main classicals can extend into the outer classical belt: 
if present, a certain number of detections of this population would be made by \textsc{ossos}.
We used the same population $P(a) \propto a^{-2.5}$ distribution as in \citet{Petit:2011p3938} (Appendix A) and as in \S~\ref{sec:classical}. 
The $q$ distribution of the component was allowed to be wider than in the L7 model, 
going from 38~\au to the $a$ value being tested.
We excluded component $a$ values that occurred in the 47.4--48.2 region occupied by the 2:1 resonance. 
Using the \ossos survey simulator, we confirm that the detection of one low-$i$, $a > 47$~\au object, 
as we found in this survey (\texttt{o3e45}),
is consistent with a stirred component smoothly extending to at least 49~\textsc{au}. This model could not be rejected by the detections.
The further the stirred component extends, the higher the number of low-$i$, $a > 47$~\au detections that should be made by \textsc{ossos}.
Extending this component further to 60~\au would imply 5 low-$i$, $a > 47$~\au detections by \textsc{ossos}. 
(This continues to hold, though at the 92\% confidence level, if the test is instead made with the power law of the distribution steepened up to $P(a) \propto a^{-4.5}$).
As we have only one such \ossos detection in the sample presented in this work, we reject a stirred component extending beyond 60~\au at the 95\% confidence level, under the assumption that the smooth extension is a power law.

An alternate hypothesis is that \texttt{o3e45} and the three previously discovered low-$i$, $a > 47$~\au objects are 
simply the low-$i$ tail of the hot population of the main Kuiper belt. 
We find that \texttt{o3e45} has a low probability of being a member of this hot component.
The \cfeps L7 model predicts that the 13A \ossos blocks have just 5\% probability of detecting 
one or more hot component objects in the $a > 47.5$~\textsc{au}, $i < 5^{\circ}$, $q > 40$ zone, where we have one detection.  
Detection of three to four more objects in this zone of orbit parameter space is needed 
before more conclusive statements can be made, 
to determine the abundance of such objects in future \ossos blocks relative to the abundance of the hot population.

\section{Conclusion}

We report 85 trans-Neptunian objects found in two distinct 21-deg$^{2}$ blocks of sky, 
monitored in the first quarter of the Outer Solar System Origins Survey (\textsc{ossos}).
These \tnos\ were discovered in 2013 and tracked through 2013--2014 with CFHT's MegaPrime wide-field square-degree imager. 
They comprise 1 Centaur, 39 resonant objects, 37 classical objects, 2 detached objects, and 6 scattering objects.

This sample is without ephemeris bias, as it is 100\% tracked above the characterization magnitude, a first for large surveys of the Kuiper belt. The orbital elements of the discoveries are precise to at least $\sigma_{a} < 1\%$, with most having $\sigma_{a} < 0.1\%$ after 12--17-month arcs. This accuracy was achieved in a significantly shorter period than in most previous surveys, thanks to the internally consistent astrometric catalogue and increased observing cadence. These 85 objects, together with their precisely quantified detection biases, can immediately be folded into the known objects usable for testing models of Solar System architecture evolution, via our survey simulator. 

This initial \ossos detected sample confirms the existence of substructure within the main classical Kuiper belt, as first reported in \citet{Petit:2011p3938}. We find that the semi-major axis distribution of the cold classicals cannot have a smooth distribution: it must contain a clumped `kernel' and a extended 'stirred' component. There is a tail of the ``stirred'' component out beyond the 2:1 resonance that extends to at least 50~\textsc{au}. Its extent beyond that is as yet unclear.

\acknowledgments
{\it Facilities:} \facility{CFHT (MegaPrime)}.

This research was supported by funding from the National Research Council of Canada and the National Science and Engineering Research Council of Canada. 
The authors recognize and acknowledge the sacred nature of Maunakea, and appreciate the opportunity to observe from the mountain. 
This project could not have been a success without the dedicated staff of the Canada--France--Hawaii Telescope (CFHT) telescope. 
Based on observations obtained with MegaPrime/MegaCam, a joint project of CFHT and CEA/DAPNIA.
CFHT is operated by the National Research Council of Canada, the Institute National des Sciences de l'Universe of the Centre National de la Recherche Scientifique of France, and the University of Hawaii, 
with this project receiving additional access due to contributions from the Institute of Astronomy and Astrophysics, Academia Sinica, National
Tsing Hua University, and National Science Council, Taiwan.
This work is based in part on data produced and hosted at the Canadian Astronomy Data Centre.
MES is supported in part by an Academia Sinica Postdoctoral Fellowship.
This research has made use of NASA's Astrophysics Data System, GNU \texttt{parallel} \citep{Tange2011a}, and many Python packages, particularly \texttt{astropy} \citep{TheAstropyCollaboration:2013cd}, \texttt{matplotlib} \citep{Hunter:2007} and \texttt{SciPy} \citep{jones:2001}; we thank their contributors for their open-source efforts.

\section{Availability}
\label{sec:code}

The development and source code are available for use and contribution from GitHub: the data pipeline at \url{https://github.com/OSSOS/MOP}, and the survey simulator at \url{https://github.com/OSSOS/SurveySimulator}. 

\section{Appendix A}
\label{sec:appendixSS}

Contrary to deep stellar or galactic surveys which analyze stacked images,
moving object surveys rely on detecting the source on each and every single
image of the discovery triplet.
For a given intrinsic magnitude, an object can appear brighter or dimmer due to
Poisson fluctuations of the source itself and of the background. Thus, the
measured magnitude scatters around the intrinsic value. For objects much
brighter than the detection limit, the scatter is small in relative value, but it
becomes important close to the limit.
This scatter produces an asymmetry in the magnitude of measured objects: objects whose
magnitudes scatter up will be easier to detect and preferentially retained,
while those that scatter down will be too dim to be detected (Malmquist bias).
This effect can be seen in Fig.~\ref{fig:trombscat}. At the faint end, we
clearly see the asymmetry with more objects having a lower apparent magnitude,
i.e. brighter, than the intrinsic magnitude.

For the \ossos simulator, the statistics of measured apparent magnitude versus intrinsic magnitude
determined here also allows us to simulate the scatter and apply it to the
intrinsic magnitude of the model objects to obtain a simulated measured
magnitude. This is the magnitude that will be used to compare with the real
detections.
To decrease the RMS of the magnitude uncertainty, creating less noise in the
determination of the slope and consequently on the population estimate error,
we took the mean magnitude of the object on the discovery triplet as the
defining magnitude of the object that we place in the simulator for comparison to the simulated detections. 
If one or more of the triplet's sources was not appropriate for photometry, e.g. due to involvement
with a star or galaxy, we excluded it from this mean. Out of
the 85 objects in the characterized sample from the 13AE and 13AO blocks, 2
objects had only one useful photometric measurement and 12 objects had only two.
For each simulated detection, we determine the mean and standard deviation of
the magnitude scatter, following the trends determined on the fake implanted
objects (see Fig.~\ref{fig:trombone}), and draw a Gaussian distributed random
number with these parameters. This yields a simulated measured magnitude. We
repeat this procedure 1, 2 or 3 times following the frequency determined on the
real/fake detections. We finally average the simulated measured magnitudes to
obtain the surmised magnitude which will be compared to the average magnitude
of the real detections.

\begin{figure}
\plotone{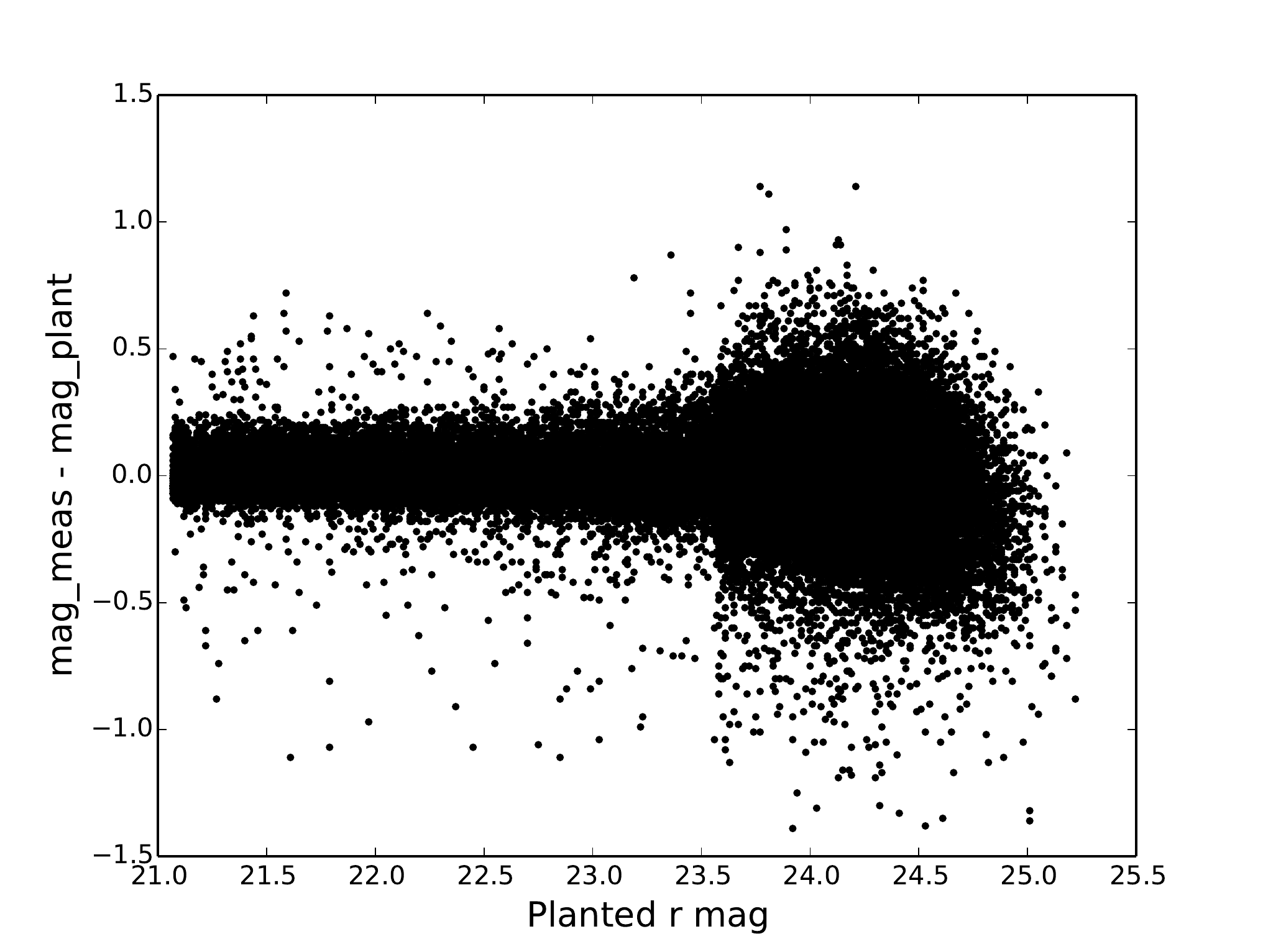}
\caption{Difference of measured magnitude from intrinsic magnitude for all fake objects
  implanted in the 13AO block to determine the detection efficiency.
  We plant a greater density of objects $m_r > 23.5$ to ensure that this magnitude range where the
  detection efficiency decreases is well quantified.
\label{fig:trombscat}}
\end{figure}

\begin{figure}
\plotone{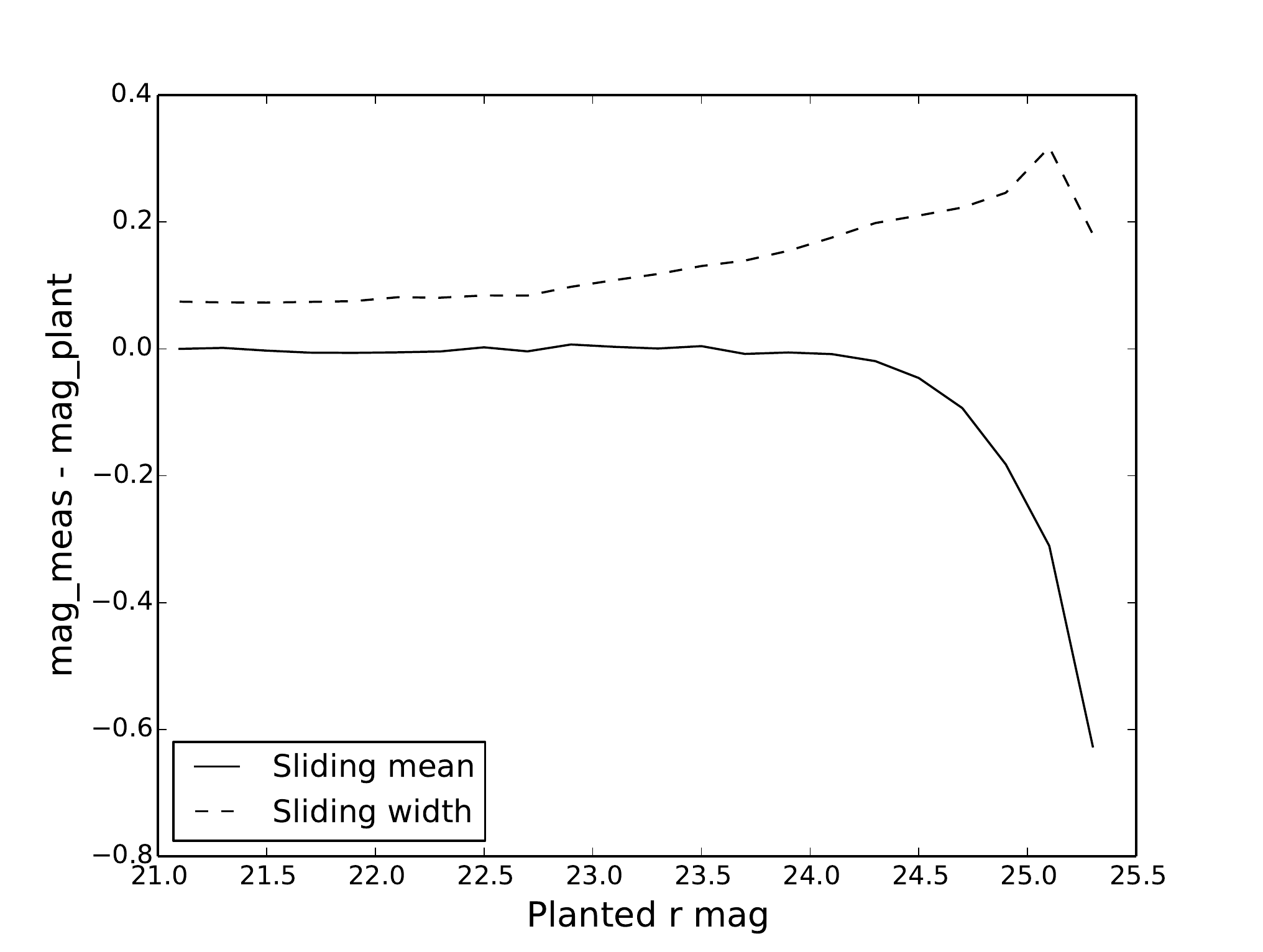}
\caption{Mean (solid line) and standard deviation (dashed line) of measured
  magnitude minus intrinsic magnitude versus intrinsic magnitudes for all
  implanted objects in the 13AO block. Mean and standard deviation have been
  evaluated on 0.2 magnitude bins, with iterative rejection of points beyond
  3-$\sigma$.
\label{fig:trombone}}
\end{figure}


\section{Appendix B}
\label{sec:pointings}


\begin{deluxetable}{crrrrlcl}
\tablecolumns{7}
\tablecaption{Area units comprising the sky observed for discovery in 13A by \textsc{ossos}, as used in the survey simulator \label{tab:pointings}}
\tablehead{\colhead{Width ($^{\circ}$)} & \colhead{Height ($^{\circ}$)} & \colhead{RA (hrs)} & \colhead{Dec ($^{\circ}$)} & \colhead{Epoch (JD)} & \colhead{Filling factor} & \colhead{Block}
}
\startdata
3.000 & 3.000 & 14:24:05.80 & -13:15:45.9 & 2456386.95496 & 0.9079 & 2013AE \\
1.000 & 3.000 & 14:15:54.20 & -12:34:03.0 & 2456386.95496 & 0.9079 & 2013AE \\
3.000 & 3.000 & 14:07:42.60 & -11:52:20.1 & 2456386.95496 & 0.9079 & 2013AE \\
3.000 & 3.000 & 16:06:13.18 & -12:43:40.5 & 2456420.95956 & 0.9055 & 2013AO \\
1.000 & 3.000 & 15:58:01.35 & -12:19:54.2 & 2456420.95956 & 0.9055 & 2013AO \\
3.000 & 3.000 & 15:49:49.52 & -11:56:07.9 & 2456420.95956 & 0.9055 & 2013AO \\
\enddata
\tablecomments{When testing the visibility of model objects due to their location on the sky, we apply a filling factor correction:  
91.64\% of the region inside the outer boundary of the MegaPrime mosaic (\S~\ref{sec:parameters}) is active pixels. 
We fold into that correction the rare pipeline failures that left a few of the 756 CCDs in each block un-searched (99.07\% success on 13AE and 98.81\% on 13AO). 
The survey simulator uses a single date for each block, 
as that is statistically equivalent (the simulator produces a statistical ensemble that is representative of the detections),  
and this approximation provides computational efficiency.
}
\end{deluxetable}

\bibliography{firstquarter}
\bibliographystyle{apj}

\end{document}